\documentclass[twocolumn]{pasj00}
\usepackage{lscape}
\renewcommand{\tabcolsep}{3pt}

\begin{document}

\title{ Near infrared spectroscopy of M dwarfs. III.\\
Carbon and oxygen abundances in late M dwarfs including the dusty 
rapid rotator 2MASSI J1835379+325954
\thanks{Based on
data collected at Subaru Telescope, which is operated by the National
Astronomical Observatory of Japan.}}

%%% begin:list of authors
% Do NOT capitalize all letters in "textsc".
\author{Takashi \textsc{Tsuji}} %
%  \thanks{Example: Present Address is xxxxxxxxxx}}
\affil{Institute of Astronomy, School of Science, The University of Tokyo,
2-21-1 Osawa, Mitaka-shi, Tokyo, 181-0015}
\email{ttsuji@ioa.s.u-tokyo.ac.jp }

\and

\author{Tadashi \textsc{Nakajima}}
\affil{National Astronomical Observatory of Japan, 2-21-1 Osawa,
Mitaka-shi, Tokyo, 181-8588}
\email{tadashi.nakajima@nao.ac.jp}


\maketitle

\begin{abstract}

Carbon and oxygen abundances of eight late M dwarfs are determined 
based on the near infrared spectra of medium resolution ($ R = 
\lambda/\Delta\,\lambda \approx 20000 $).
In late M dwarfs,  dust forms in their photospheres for $T_{\rm eff}$ 
below about 2600\,K, and this case applies to the M8.5 dwarf 2MASSI 
J1835379+325954 (hereafter 2MASS 1835+32) whose $T_{\rm eff}$ is 2275\,K. 
Other seven  objects with $T_{\rm eff}$ above 2600\,K are analyzed with the 
dust-free models.
For the case of 2MASS 1835+32 analyzed by the dusty model, the surface 
temperature is higher by about 600\,K due to the blanketing effect of the 
dust grains, mainly composed of iron grains, 
and the carbon and oxygen abundances are higher by
about 0.25 and 0.15 dex, respectively, compared to the analysis by
the dust-free model. Once dust forms in the
photosphere, the dust works as a kind of thermostat and temperatures
of the surface layers remain nearly the same as the condensation temperatures
of the dust grains. For this reason, the temperatures of the
surface layers of the dusty dwarfs are relatively insensitive to the
fundamental parameters including $T_{\rm eff}$. In addition, it appears that
2MASS 1835+32 is a rapid rotator, for which its EWs are thought to 
remain unchanged by the rotational broadening. This is, however,
true only when the true continuum is well defined. Otherwise, the
pseudo-continuum level depends on the rotational velocity and hence the EWs
as well. For this reason, the derived abundances depend on the rotational
velocity assumed: For the values of $V_{\rm rot}$sin\,$i$ = 37.6 and 
44.0\,km\,s$^{-1}$ available in the literature, the derived carbon and
oxygen abundances differ by  0.23 and 0.14\,dex, respectively, and we find 
that the higher value provides a better account of the observed spectrum. 
The resulting carbon and oxygen abundances in the eight late M dwarfs show
no systematic difference from our results for the early and middle M dwarfs,
and confirm the higher $A_{\rm O}/A_{\rm C}$ ratio at the lower metallicity.
In late M dwarfs, CO and H$_2$O remain as excellent abundance indicators 
of carbon and oxygen, respectively, except for additional uncertainty due to 
the complexity associated with the dust formation in the latest M dwarfs.
\end{abstract}

\section{Introduction}

The late M dwarfs are the last stars that shine by their own nuclear burning 
in their interiors and mark the end of the main sequence. It was shown 
that the lower limit of the stellar mass that maintains the high enough 
central temperature to ignite hydrogen burning is 0.08\,$M_{\odot}$
(\cite{Hay63}; \cite{Kum63}). In recent years, more attention is  
directed to the objects below the hydrogen-burning limit,
referred to as brown dwarfs, rather than to the
lowest mass hydrogen-burning objects, the latest M dwarfs.    
At the end of the 20-th century, extensive searches for a brown dwarf
have been attempted by several groups with different methods, and a genuine 
brown dwarf, Gliese\,229B, was finally discovered by \citet{Nak95} with 
immediate spectroscopic verification by \citet{Opp95}. Since then, 
many brown dwarfs were discovered by the large scale surveys such as
DENIS \citep{Epc99}, 2MASS \citep{Skr06}, SDSS \citep{Yor00}, UKIDSS
\citep{Law07}, and WISE \citep{Wri10}.

In observations, however, it is by no means easy to determine the stellar mass 
accurately, and hence it is not very sure if some latest M dwarfs are really 
stars or young brown dwarfs. Anyhow, some of our objects are situated 
at the boundary between  stars and  brown dwarfs, and we hope that a
detailed spectroscopic study of such objects at relatively high 
resolution would be of some use in extending  detailed spectroscopic 
analyses to very low mass objects. In fact,
the high resolution spectroscopy was already applied to a 
cool brown dwarf  more than a decade ago by \citet{Smi03}, who 
observed the nearby T dwarf $\epsilon$ Indi Ba with the 
resolution as high as $R \approx 50000 $ and showed the advantages of
the high resolution spectroscopy over the other works mostly based on 
low resolutions.
Thus, the very cool dwarfs including  brown dwarfs are  already within
the capability of the present-day high resolution infrared spectroscopy
and we are even too late to work on M dwarfs.
  
Now, returning to  M dwarfs, one problem in the  latest M dwarfs is 
that dust may form in their photospheres. In fact,
it was shown that the thermochemical
conditions for the formation of dust grains such as corundum, iron, and
enstatite are well met in the photospheres of M dwarfs with $T_{\rm eff}$
below about 2600\,K, and  actual model photospheres in thermal and convective
equilibria incorporating the dust formation have been generated  
\citep{Tsu96a}. The predicted
spectra based on the dusty models appeared to be consistent with the
known observation of the infrared spectrum of the latest M dwarf
LHS\,2924, while the dust-free models predicted H$_2$O bands to be
too deep compared with the observed spectrum. 

The evidence for  dust has more clearly been shown in a peculiar
cool dwarf GD\,165B discovered by \citet{Bec88}. This object 
showed a very red color
but molecular bands were not so strong as expected for a possible 
low temperature suggested by the very red color. Such a characteristic
could have been explained nicely by our dusty model \citep{Tsu96b}:
The very red color is due to infrared excess caused by dust,
which also weakens the molecular absorption bands at the same time.
This object was later recognized as a brown dwarf and regarded as the
proto-type of L type dwarfs \citep{Kir99}.  Also an analysis of  optical
spectra has shown additional spectroscopic evidence for dust 
in late M dwarfs \citep{Jon97}. Since then the problem of
the dusty photospheres has been studied mostly in connection with 
brown dwarfs both observationally and theoretically (e.g., see an extensive
survey of such efforts by \cite{Hel08a}). More recently, modeling of the
cool dwarfs has made significant progress, extending to the cooler 
brown dwarfs and extra-solar planets (see reviews, e.g., \cite{Hel14};
 \cite{Mar15}) on one hand, and also to self-consistent cloudy models 
of late M and L dwarfs \citep{Wit11} on the other.

In addition to the problem of dust, some of the late M dwarfs
are known to be rapid rotators. One of our targets  2MASS 1835+32,
which was recognized as a nearby late-type
M dwarf only recently \citep{Rei03}, is also found to be
a rapid rotator (\cite{Rei10}; \cite{Des12}).
In late M dwarfs in which molecular lines are already blended each other,
the rapid rotation further smears out the blended spectra.  We have
developed a method to analyze the spectrum without the true-continuum by
referring to the pseudo-continuum which is depressed appreciably by
 molecular bands (\cite{Tsu14}; \cite{Tsu15}, hereafter
Papers I and II, respectively). We find that our method can be extended
quite well to the case of the pseudo-continuum smeared-out further by the 
rapid rotation, and also can be applied to determine the rotational 
velocity more consistently.
  
In this paper, we first introduce our observed data (section 2) and then
we discuss fundamental parameters such as effective temperatures of
late M dwarfs (section 3). We then generate model photospheres 
including the dusty case, and examine the effect of dust  on the 
thermal structures and spectra with our simple dusty models (section 4). 
We first  analyze seven non-rapid rotators,
whose effective temperatures are found to be above 2600\,K and hence can be
assumed to be dust-free (section 5).
Then, we examine the  rotational broadening on the spectrum of 2MASS 1835+32
(section 6). We examine the effects of  dust formation, rotational velocity, 
and fundamental parameters on our abundance analysis (section 7).
 Finally, we discuss our results on the carbon and oxygen abundances in 
M dwarfs, dust in late M dwarfs, and the coolest end of the main
sequence (section 8).

\section{Observations}

We observed eight late M dwarfs between M4.5 and M8.5 listed in table 1.
Observations were carried out at the Subaru Telescope on 2014 August 31 
(UT) using the
echelle mode of the Infrared Camera and Spectrograph (IRCS) 
\citep{Kob00} with adaptive optics.
Six of the eight targets were bright enough in the visible to be natural 
guide stars, while LP 412-31 
and 2MASS 1835+32 were too faint and laser guide stars were used.
The slit width of 0.14$^{\prime\prime}$ was sampled at 55 mas pixel$^{-1}$, 
and the resolution 
was about 20000 at $K$. The echelle setting was ``$K^+$'', which covered 
about a half of the $K$ window with the orders, wavelength segments, 
and pixel-wise dispersions given in table 2.   
The targets were nodded along the slit, and observations were taken in an 
ABBA sequence, where A and B
stand for the first and second positions on the slit.
The total exposure time ranged from 24s (GJ 873) to 80 min (LP 412-31).
The night was photometric at the beginning, but the condition 
gradually degraded and it was spectroscopic toward the end of the night. 
The visible seeing was better than 0.5$^{\prime\prime}$ throughout the night.
Near the mid night, a rapidly rotating B8V star, HR 326 was observed as 
the calibrator of telluric transmission. Signal-to-noise ratios of reduced 
spectra are given for the 28th and 25th orders in table 1. 
H$_2$O and CO lines are analyzed  in the 28th  order (aperture 2), and
 H$_2$O lines are also analyzed in the 25th order (aperture 5). 

 Data reduction was carried out using the standard IRAF\footnote{IRAF is 
distributed by the
 National Optical Astronomy Observatory, which is operated by the 
Association of Universities
 for Research in Astronomy, Inc., under cooperative agreement with the 
National Science Foundation.}
 routines in the {\texttt{imred} and \texttt{echelle} packages. 
 After extraction of one-dimensional spectra, wavelength calibrations 
were calculated using telluric
 absorption lines in the spectra of HR 326. After wavelength calibrations 
of one-dimensional spectra
 of A and B positions, they were co-added to produce combined spectra. 
The combined spectra were
 normalized by the pseudo-continuum levels and then calibrated for 
telluric absorption using the  spectra of HR 326.  

\vspace{2mm}

--------------------

table 1: p.29

table 2: p.29

--------------------

\section{Fundamental parameters}
Angular diameter measurement by interferometry is not yet extended to
late M dwarfs, and we  apply the $M_{\rm 3.4} - {\rm log}\,T_{\rm eff}$
($M_{\rm 3.4}$ is the absolute magnitude at 3.4\,$\mu$m based on the
WISE data \citep{Wri10}) relation introduced in Paper I, with a 
slight modification for the latest M dwarfs.

\subsection{Effective temperatures}

For the M dwarfs with $M_{3.4} \lesssim 9.0$ (dM4.5 $-$ dM6) in our present 
sample, effective temperatures are estimated by the use of the
$ M_{\rm 3.4} - {\rm log}\,T_{\rm eff} $ relation shown by the dashed line 
on figure 1 of Paper I ( a part of which is reproduced in figure 1a by 
a dashed line). This method works well for these M dwarfs and the results for 
four objects, GJ\,54.1, 873, 1002, and 1245B, are given in table 3.

For the later M dwarfs of $M_{3.4} > 9.0$ (dM7 $-$ dM8.5), however,
a dashed line (figure 1a) used to estimate $T_{\rm eff} $  was defined 
only by four objects, GJ\,406, 644C, 752B and 3849, whose effective 
temperatures were determined by the infrared flux method \citep{Tsu96a}. We
estimate the physical parameters of these four  objects in table 4
and examine the results in comparison with those of the evolutionary models
by \citet{Bar98}. 
As shown in  the lower right corner of the HR diagram in figure\,17 of 
Paper I and reproduced in figure\,1b, three of these four objects 
(GJ\,406, 644C, and 3849) agree well with the theoretical 
HR diagram by \citet{Bar98} shown by a solid line. However,  one 
object (GJ\,752B) deviates appreciably from the theoretical HR diagram.
For this reason, we adopt these three objects (GJ\,406, 644C, and 3849)
as reliable calibration objects, and use these three objects to define
a revised $ M_{\rm 3.4} - {\rm log}\,T_{\rm eff} $ relation shown by
a solid line in figure\,1a. The resulting values of $T_{\rm eff} $ based on 
the revised $ M_{\rm 3.4} - {\rm log}\,T_{\rm eff} $ relation are given 
in table 3 for GJ\,752B, GAT\,1370, LP\,412-31, and 2MASS\,1835+32.

\vspace{2mm}

--------------------

figure 1: p.15

\vspace{2mm}

table 3: p.29

table 4: p.30

--------------------

\subsection{Surface gravities}
For estimating the surface gravity, radius and mass are needed. 
For this purpose, 
 the effective temperature vs. radius relation given by equation 8 
and the mass vs. radius relation given by equation 10 of \citet{Boy12}
are applied to the dM4.5 - dM6 dwarfs and the results are given in table 3
together with the resulting values of log\,$g$.

For the later dM7 - dM8.5 dwarfs, we try to extend the relations of
\citet{Boy12} by the use of our calibration stars in table 4.
For this purpose, the mass is estimated with the use of the mass-luminosity 
($M_{\rm K}$) relation by \citet{Del00} 
extended to the lower masses ($M \lesssim 0.1 M_{\odot}$) with the use of 
the theoretical mass - $M_{\rm K}$ relation by \citet{Bar98}, as shown 
in figure\,2. This extension should be reasonable, since the empirical 
(dashed line) and theoretical (dotted line) mass-luminosity relations 
agree very well for $M >0.1 M_{\odot}$ in figure\,2.

With the observed bolometric flux based on the photometry, the bolometric
luminosity is derived with the known parallax, and with the $T_{\rm eff}$
based on the infrared flux method, the radius is estimated. 
 The radius vs. effective temperature relation given by equation 8 
of \citet{Boy12} shown by a dashed line in figure\,3 is extended to the 
lower temperatures with our calibration stars nos. 1, 2, and 4 
in table 4, as shown by a solid line in figure\,3. It is to be noted that
our empirical data for three calibration stars agree rather well with the
theoretical relation based on the evolutionary models of the solar 
metallicity \citep{Bar98} shown by a dotted line in figure 3.

The radius vs. mass relation given by equation 10 of \citet{Boy12} shown by 
a dashed line in figure 4 is extended to the lower masses 
($M \lesssim 0.1 M_{\odot}$) 
with the use of the calibration stars nos.\,1, 2, and 4 in table\,4, 
as shown by a solid line in figure\,4. The dashed and solid lines are
joined at $ M \approx 0.15\,M_{\odot}$. For comparison, predicted relation by
the evolutionary models of the solar metallicity \citep{Bar98}
is shown by a dotted line in figure.4. Again the empirical relation based 
on our three calibration objects agrees rather well with the theoretical 
result of \citet{Bar98}. 

Now, radii and masses of our late M dwarfs  are estimated with the use of 
the solid lines in figures\,3 and 4, respectively, and the results for 
GJ\,752B, GAT\,1370, LP\,412-31, and 2MASS\,1835+32 are given in
table\,3 together with the resulting values of log\,$g$. 

\vspace{2mm}

--------------------

figure 2: p.15

figure 3: p.15

figure 4: p.15

--------------------

\section{Model photospheres of late M dwarfs}
  We apply dust-free  model photospheres for the M dwarfs 
with $T_{\rm eff} \gtrsim 2600$\,K (subsection 4.1). Since our sample
now includes the latest M dwarf in which dust may form, we
examine dusty models in some detail (subsection 4.2), and effects of
dust on the thermal structures (subsection 4.3) and spectra (subsection 4.4)
are discussed.  

\subsection{Dust-free model photospheres of M dwarfs}

The $T_{\rm eff} $ values of  
our seven program stars are higher than 2600\,K (table 3), and
we apply the dust-free models of case C to these seven objects.
The case C includes Ca series based on  the abundance {\it case a} 
(see Table 1 of \cite{Tsu02}) adopting the classical solar C and O abundances
\citep{And89} and Cc series on the abundance {\it case c} adopting the 
downward revised C and O abundances \citep{All02}. 
In our first iteration on CO spectra, we use the models
of the Ca series from our UCM (Unified Cloudy Model) 
database \footnote{UCM is a kind of semi-empirical cloudy model, but our 
UCM database includes dust-free models (case C) as well as fully dusty 
models (case B) as its subsets. As for detail:
http://www.mtk.ioa.s.u-tokyo.ac.jp/$\sim$ttsuji/export/ucm2.}.  
  For further iterations, we generate specified 
model for each object based on $T_{\rm eff} $ and log\,$g$ given in 
table 3, and  the models are either Ca or Cc series depending  on the carbon 
abundance determined by the first iteration on CO (section\,5).  
These models are designated as Ca or Cc/$T_{\rm eff} $/log\,$g$: for 
example, Ca3160c508 implies 
a model of Ca series, $T_{\rm eff}$ = 3160\,K, and log\,$g$ = 5.08. 
The micro-turbulent velocity is kept to be 1 km\,s$^{-1}$ throughout.

\subsection{Dusty model photospheres of the latest M dwarfs}

Dust may form in the photosphere of M dwarfs with 
$T_{\rm eff} \lesssim 2600$\,K \citep{Tsu02} 
and, since the $T_{\rm eff}$ value of our object 2MASS\,1835+32 is
found to be 2275\,K (table\,3), we consider the effect of
dust in our analysis of this M8.5 dwarf. 
In late M dwarfs, however, dust formation has just started and
only small amount of dust grains should be formed. For this reason,
we  assume the simplest LTE (local thermodynamical equilibrium) model 
that dust forms everywhere so long as 
temperature is lower than the condensation temperature, $ T_{\rm cond}$
($T \lesssim T_{\rm cond}$), and we referred to such a model as 
a fully dusty model of case B \citep{Tsu02}.
However, such a fully dusty model of case B is most 
difficult  from the viewpoint of generating non-grey convective models,
possibly because very large dust opacities  compared with the
gaseous opacities dominate throughout the upper layers. 
For this reason, we could not complete a grid for case B before, but 
computed only one sequence of log\,$g$ = 5.0 \citep{Tsu02}. We now need to
generate some models of case B for the analysis of our target star
and we consider  this problem again in this section.  

We are using a simple iterative procedure by correcting the temperature
gradient by
$$ { \big({ dT \over d\tau} \big)_{\rm rev}}  =
{{ \sigma\,T_{\rm eff}^{4} -\pi\,F_{\rm conv}(\tau) } \over 
{ \pi\,F_{\rm rad}(\tau) } }
{ \big( {dT \over d\tau} \big) },   \eqno(1) $$
where $ F_{\rm rad}(\tau) $, $ F_{\rm conv}(\tau) $, and  $ dT/d\tau $ 
are the total radiative flux, convective flux based on the
local mixing-length theory (e.g., \cite{Hen65}), and temperature gradient 
of a model at hand, respectively. Then, the revised temperature gradient 
$ (dT/d\tau)_{\rm rev} $ is integrated with several trial surface 
temperatures at log\,$\tau_0 = -7.0$ ($\tau_0$ is the optical
depth defined by the continuous opacity at $\lambda = 0.81\,\mu$m),
 and a revised $T(\tau)$ is fixed so that it gives the
correct surface flux of $\pi\,F_{\rm rad}(\tau = 0) = \sigma\,T_{\rm eff}^{4} $
\citep{Tsu66}. This method was extended from the case neglecting convection,
 i.e., $F_{\rm conv}(\tau) = 0 $
\citep{Tsu65} and has an obvious meaning: The temperature gradient should be
increased (decreased) if the total flux is smaller (larger) than the 
equilibrium value of $\sigma\,T_{\rm eff}^{4} $. If convection is
included, temperature gradient is tempered to be consistent with
the smaller radiative flux, according to equation 1.

If dust forms, such an iterative procedure turns out to be unstable, possibly
because very large dust opacities are highly sensitive to the changes of the 
physical conditions in general and also appear suddenly at $T = T_{\rm cond}$.
In such a case, we find that it is useful to temper the variation of 
$F_{\rm rad}(\tau) $ somewhat. We prepare a small grid of the fully dusty 
models of case B to be applied to late M dwarfs; for $T_{\rm eff}$
between 2000 and 2600\,K, log\,$g$ = 5.0, 5.25, and 5.5, and abundances
of {\it case a} and {\it case c}. 
The results are added to our UCM database $^{2}$.
We generally aim at a flux constancy within 1\,\%, but flux errors as large
as 3\,\% remain in cooler models near $T_{\rm eff} = 2000$\,K in 
case B \footnote{The convergence in the fully dusty models of case B for
$T_{\rm eff} < 2000$\,K is more difficult as shown for these models in a 
sequence of log\,g = 5.0 so far computed$^{2}$. The models of case B, however,
are no longer useful for $T_{\rm eff} < 2000$\,K and, instead,  
the cloudy models should be used for L and T dwarfs.}.

\subsection{Effect of dust on the thermal structures}
In figure\,5, the thermal structures of the fully dusty models  
(blue or black) are compared with those of the dust-free models 
(light sky or grey) for  $ T_{\rm eff} = 2600$, 
2300, and 2000\,K (abundance {\it case c} and log\,$g$ = 5.25). 
The condensation lines of corundum (Al$_2$O$_3$), iron (Fe) and
enstatite (MgSiO$_3$) are shown by dashed lines.  The dust-free models
penetrate to the regime of $ T \lesssim T_{\rm cond} $ in the
surface layers and the thermodynamical condition for dust formation
is fulfilled in all the models shown in figure\,5. Thus corundum first 
forms and temperatures are elevated
somewhat by its blanketing effect. Then iron forms, and  because of the 
larger abundance of iron, it has dominant effect: the temperatures of 
the upper layers coincide with the condensation line of iron as shown in 
figure\,5. This is a result that the photosphere is forced to be in 
radiative equilibrium, and it seems that  iron  works as a kind of
thermostat; if a temperature $T_{\rm f}$ within  the 
condensation line of iron in our model
is increased by  perturbation, then the iron grains will evaporate and
the temperature will decrease because the blanketing effect of the
iron grains disappears. But, if the temperature goes down below $T_{\rm f}$,
iron grains form and the temperature will increase again because 
of the blanketing effect of the iron grains. Such processes repeat and
the temperature will eventually settle at
 the  starting value $T_{\rm f}$. As a result, the temperatures converge to 
the condensation line of iron and the thermal 
structures of the dusty models in the surface layers are not much
different for late M dwarfs of different effective temperatures, in marked
contrast to the dust-free models in which surface temperatures
show large differences for different effective temperatures, 
as shown in figure 5.
It is to be noted, however, that enstatite could not be formed
since the photospheres are warmer than the condensation temperatures
of enstatite because of the blanketing effect of  iron and corundum.

\vspace{2mm}

--------------------

figure 5: p.16

--------------------

\subsection{Effect of dust on the spectra}
The predicted spectra based on the dusty and dust-free models
discussed in the previous subsection are shown in figure 6. 
In the model spectra of $T_{\rm eff}$ = 2600\,K shown in the top of 
figure\,6, the difference of the dusty (shown by blue or black) and
dust-free (light sky or grey) cases is rather minor: the dusty model shows
slightly weaker absorption only in the regions of strong absorption 
(e.g., at 2.5\,$\mu$m) because of its higher surface temperature. 
Although the temperatures of the surface region are about
400\,K higher in the dusty model than in the dust-free model (see figure\,5),
the matter density in the surface region is very low and this
region has little effect on the emergent spectra except for the very
strong absorption formed in the very surface. For this reason, we
neglect the effect of dust for M dwarfs with $T_{\rm eff} \approx$ 2600\,K
(e.g., GJ\,752B, LP\,412-31).

In the model spectra of $T_{\rm eff}$ = 2300\,K shown in the middle of 
figure\,6, the effect of dust is considerable, since the
larger part of the surface region is now warmer (by about 600\,K) 
in the dusty model than in the dust-free model (see figure 5).
Especially, strong bands such as H$_2$O 1.9 and 2.7\,$\mu$m  
bands as well as CO first overtone bands formed in the  surface 
layers suffer  appreciable effect, while weak lines such as at 2.2
$\mu$m region formed in the deeper layers show  minor change.
However, dust has considerable effect even on the weak lines 
on the higher resolution spectra, as will be shown in section\,7.

 In the case of $T_{\rm eff}$ = 2000\,K shown in the bottom of
figure\,6, the effect of dust is more drastic: The spectrum of
the dusty model shows overall excess compared with the dust-free
model in the $K$ band region and this is of course compensated for
by the large extinction due to dust in the shorter wavelength region.
The very strong absorption bands due to H$_2$O and CO in the dust-free
model are much weaker in the dusty model because of the elevated 
temperatures in the larger part of the photosphere and also by 
the dust extinction. But $T_{\rm eff}$ = 2000\,K is already in the 
regime of L dwarfs, and late M dwarfs are still free 
from such drastic effect of dust.

\vspace{2mm}

--------------------

figure 6: p.16

--------------------

\section{Seven M dwarfs of  non-rapid rotators}
Seven M dwarfs in our sample except for 2MASS\,1835+32 are not
rapid rotators and also the effect of dust may not be important
($T_{\rm eff} > 2600$\,K). Then, CO and H$_2$O lines are analyzed by 
the mini curves-of-growth (subsection 5.1) and by the synthetic spectra 
(subsection 5.2), as in Papers I and II.

\subsection{The mini curve-of-growth analysis }

\subsubsection{First iteration on the CO blends}
We analyze  the CO blends of the 2-0 band listed in  table 7 of Paper I with
the solar carbon abundance of {\it case a} as an initial
starting value for each object. Equivalent widths (EWs)
are measured by referring to the pseudo-continuum and the
resulting values of log\,$W/\lambda$ are given in table 5.
We then apply the mini curve-of-growth (CG) method 
from the beginning (i.e., not apply the conventional analysis as used 
in the first iteration on CO in Paper I).

The mini CG method was developed through our Papers I to II: 
Briefly summarizing, we first  generate synthetic spectra for an assumed
initial abundance log\,$A_0$ and  corrected  abundances
log\,$A_0 \pm \delta$. Then, equivalent widths, $W$'s, are evaluated from 
these synthetic spectra by the same way as the observed EWs are measured, 
referring to the pseudo-continuum  which is evaluated accurately 
by the use of the recent line-list of H$_2$O (\cite{Bar06}, \cite{Rot10}). 
In the first iteration, we apply the models from the UCM grid and assume 
$\delta = 0.3$. Then we measure equivalent widths, $W$'s, on the 
predicted spectra calculated with the starting initial values of
log\,$A_{\rm C} = -3.40$ \footnote{We use the notation: $A_{\rm El} =
N_{\rm El}/N_{\rm H} $, where $N_{\rm El}$ and $N_{\rm H}$ are the number 
density of the element El and hydrogen, respectively.} 
 and log\,$A_{\rm O} = -3.08$ ({\it case a} for the 
initial model of Ca series) and with log\,$A_{\rm C} = -3.40 \pm \delta$ and 
log\,$A_{\rm O} = -3.08 \pm \delta$ ($\delta$ = 0.3) \footnote{The CO 
abundance depends almost on the carbon abundance alone
so far as $A_{\rm C} < A_{\rm O}$. However, this may no longer be true if
$A_{\rm C} \approx A_{\rm O}$ and, to prevent such a circumstance when
carbon alone is increased, we also changed oxygen abundance so that
$A_{\rm O}/A_{\rm C}$ ratio remains the same.}. These EWs
are designated as log\,$W(\delta)/\lambda$. Then, we generate a
mini CG  plotting log\,$W(\delta)/\lambda$ against $\delta$ = -0.3, 0.0,
and +0.3 for each blend.  With this mini CG, observed EW is converted to 
the abundance correction to the assumed abundance of carbon. As for more 
details about the mini CG method, see sections 4.1  and 6.3  of Paper II.

The above noted process is repeated blend-by-blend for all the EWs of CO 
measured in table\,5. The resulting abundance corrections  to the starting
value of log $A_{\rm C}^{(0)} = -3.40 $ for all the measured EWs
given in table\,5 are shown in figures\,7a-g for our seven program stars. 
The mean abundance correction $\Delta\,{\rm log}\,A_{\rm C}^{(1)}$
and the resulting carbon abundance
log\,$A_{\rm C}^{(1)} = -3.40 + \Delta\,{\rm log}\,A_{\rm C}^{(1)}$ are 
given in the fifth and sixth columns of table 6, respectively, for each object. 

\vspace{3mm} 
\subsubsection{First iteration on the H$_2$O blends in  region B}

Since H$_2$O lines are sufficiently strong in region B
(defined in figure 3 and detailed in section\,3.4 of Paper II)
for late M dwarfs,  we first analyze
 the H$_2$O blends in  region B listed in table 4 of Paper II.
EWs measured by referring to the pseudo-continuum are given in table 7.

Given that the carbon abundance is now known to be log\,$A_{\rm C}^{(1)}$
from the analysis of CO in the preceding subsubsection, 
we apply the specified model for each object 
noted in the second column of table 8. Assuming an initial starting value of 
log\,$A_{\rm O}^{(0)} = {\rm log}\,A_{\rm C}^{(1)} + 0.30$ \footnote{
If the starting value of log\,$A_{\rm O}^{(0)}$ is fixed to be the solar 
abundance as in the case of log\,$A_{\rm C}^{(0)}$, it is possible that 
$ A_{\rm O}^{(0)} \lesssim A_{\rm C}^{(1)} $ depending on the analysis on 
carbon abundance in the preceding subsection. For this reason, we assume a
more plausible starting oxygen abundance to be $ A_{\rm O}/A_{\rm C} 
\approx 2 $ as in the Sun \citep{All02}. },
the mini CG is generated from the synthetic spectra for 
log\,$A_{\rm O}^{(0)}$ and log $A_{\rm O}^{(0)} \pm \delta$ \footnote{The 
H$_2$O abundance depends not  on the oxygen abundance, but  
on $ A_{\rm O} - A_{\rm C} $. Then, unlike the case of CO in
the preceding subsubsection, we fix the carbon abundance to be
log\,$A_{\rm C}^{(1)}$ and only the oxygen  abundance is changed to apply 
the mini CG method to the H$_2$O blends.} for each object.
Since this  starting value of oxygen may be fairly good as can be 
expected from our experience so far,  we now apply the mini CG method with 
$\delta = 0.1$ instead of 0.3. The resulting oxygen abundance corrections 
$\Delta\,{\rm log}\,A_{\rm O}^{(1)}$ are shown in figures\,8a-g.
The mean abundance correction $\Delta\,{\rm log}\,A_{\rm O}^{(1)}$
 and the resulting oxygen abundance 
log\,$A_{\rm O}^{(1)} =  ({\rm log}\,A_{\rm C}^{(1)} + 0.30) +
 \Delta\,{\rm log}\,A_{\rm O}^{(1)}$
are given in the third and fourth columns of table\,8, respectively,
for each object.  

\vspace{3mm}
\subsubsection{Second iteration on the CO blends}

Starting from log\,$A_{\rm C}^{(1)}$ and log\,$A_{\rm O}^{(1)}$
 in tables\,6 and 8, respectively, we proceed to the second iteration
on the carbon abundance, using the specified model for each object
noted in the seventh column of table 6. The resulting
second carbon abundance corrections $\Delta\,{\rm log}\,A_{\rm C}^{(2)}$
show more or less similar patterns as those in figures 7a-g. 
The second carbon abundance correction $\Delta\,{\rm log}\,A_{\rm C}^{(2)}$,
which is generally smaller than $\Delta\,{\rm log}\,A_{\rm C}^{(1)}$,
 and the resulting carbon abundance 
log\,$ A_{\rm C}^{(2)} = {\rm log}\,A_{\rm C}^{(1)} +
 \Delta\,{\rm log}\,A_{\rm C}^{(2)} $
are given in the eighth and ninth columns of table\,6, respectively,
for each object.  Although we do not illustrate the results as in 
figures 7a-g, the decrease of the probable errors in the resulting
abundances (compare columns 6 and 9 in table 6) confirms that the
accuracy of the analysis is improved by the use of the better
starting values in the second iteration.

\vspace{3mm}
\subsubsection{Second iteration on the H$_2$O blends in  region B }

Starting from log\,$A_{\rm C}^{(2)}$ and log\,$A_{\rm O}^{(1)}$ 
 in tables\,6 and 8, respectively,
second oxygen abundance corrections  $\Delta\,{\rm log}\,A_{\rm O}^{(2)}$ 
are obtained  and the results also show  more or less similar patterns
as those of figures\,8a-g.
The second oxygen abundance correction $\Delta\,{\rm log}\,A_{\rm O}^{(2)}$
and the resulting oxygen abundance 
log\,$ A_{\rm O}^{(2)} = {\rm log}\,A_{\rm O}^{(1)} +
 \Delta\,{\rm log}\,A_{\rm O}^{(2)} $
are given in the fifth and sixth  columns, respectively, of table\,8
for each object.  

\vspace{3mm}
\subsubsection{Further Iteration on the H$_2$O blends in  region A }

From our experience in Paper II, we first thought that it is sufficient to 
analyze the H$_2$O blends only in region B for late M dwarfs. However, 
it appears
that the H$_2$O lines in  region B are rather weak and noisy in the 
hottest object in our sample, GJ\,873 (see figure\,11a). Also, it appears
that the spectrum of  region B is disturbed by unknown cause
in the coolest object  LP 412-31 in our sample (see figure\,11g).   
For these reasons,  we decide to analyze the H$_2$O blends in  region A
(defined in figure 2 and detailed in section 3.3 of Paper II), 
where the H$_2$O lines are stronger. EWs are measured for
the H$_2$O blends listed in table 2 of Paper II and the results are given
in table\,9.

We start from log\,$A_{\rm C}^{(2)}$ and log\,$A_{\rm O}^{(2)}$ 
in tables\,6 and 8, respectively,
and oxygen abundance corrections  $\Delta\,{\rm log}\,A_{\rm O}^{(3)}$
are obtained.  The results are  shown in figures\,9a-g.
The oxygen abundance correction $\Delta\,{\rm log}\,A_{\rm O}^{(3)}$ and
the resulting oxygen abundance 
log\,$ A_{\rm O}^{(3)} = {\rm log}\,A_{\rm O}^{(2)} +
\Delta\,{\rm log}\,A_{\rm O}^{(3)} $
are given in the third and fourth columns, respectively, of table\,10
for each object.  
Note that the values of $\Delta\,{\rm log}\,A_{\rm O}^{(3)}$ are generally 
less than the probable errors, and this fact confirms that the resulting
oxygen abundances from region B are already well determined.
However,  the value of $\Delta\,{\rm log}\,A_{\rm O}^{(3)}$
for GJ\,873 is rather large compared with the probable error: This may be 
because the H$_2$O lines in region B are too weak in GJ\,873 and hence  
their analysis cannot be accurate as noted above.
 
\vspace{2mm}

--------------------

figure 7: p.17

figure 8: p.17

figure 9: p.18

\vspace{2mm}

table 5: p.30

table 6: p.30

table 7: p.31

table 8: p.31

table 9: p.32

table 10: p.32

--------------------

\vspace{3mm}

\subsection{Synthetic Spectra}
\vspace{1mm}
\subsubsection{CO 2-0 bandhead region}

 Based on log\,$A_{\rm C}^{(2)}$ and log\,$A_{\rm O}^{(2)}$ in tables\,6 and 
8, respectively, CO spectra near the 2-0 bandhead region are evaluated with 
the use of the specified models for our seven objects. The resulting 
predicted spectra, convolved with the slit function (Gaussian) of  FWHM =
16\,km\,s$^{-1}$, are compared with the observed ones in figures\,10a-g.
Note that the observed and predicted spectra are both normalized by
their pseudo-continua. 

The $\chi^{2}$ value  is  evaluated from
$$ \chi^{2} = {1\over{N-1}}\sum_{i=1}^{N}
{\Bigl(}{ {f_{\rm obs}^{i} - f_{\rm cal}^{i} }\over\sigma_{i}
}{\Bigr)}^{2},   \eqno(2) $$
where $f_{\rm obs}^{i}$ and  $ f_{\rm cal}^{i}$ are observed and predicted
spectra normalized by their pseudo-continua, respectively. $N$ is the number 
of data points and $\sigma_{i}$ is the noise level  estimated from the 
$S/N$ ratio in table\,1 (assumed to be independent of $i$). 
The resulting $\chi^{2}$ values are given in the last column of table\,6.
The  $\chi^{2}$ values are large for GJ\,752B and GJ\,1002, but the $S/N$
ratios for these objects are also high (table 1, fifth column). Thus, 
it is to be noted that the  $\chi^{2}$ value is by no means
a  measure of goodness of the fit. The  $\chi^{2}$ value of LP\,412-31 is 
also large even though the $S/N$ ratio is not especially high. Inspection of
figure 10g reveals that the fit is really poor in this case at the 
right edge  (hatched region). The  $\chi^{2}$ value actually plays
its expected role in this case, but this can more easily be known by the
visual inspection of the plot.

\vspace{3mm}
\subsubsection{H$_2$O in  region B}

Based on log\,$A_{\rm C}^{(2)} $ and log\,$A_{\rm O}^{(2)}$ in tables\,6 and
 8, respectively, H$_2$O spectra in a part of region B (about one fourth 
of region B) are evaluated with the use of the specified models given in 
the second column of table 8. The resulting predicted spectra are compared 
with the observed ones in figures\,11a-g.
The resulting $\chi^{2}$ values are given in the last column of table\,8.
We again find the correlation of the  $\chi^{2}$ values with the $S/N$
ratios of the observed spectra (region B is in the same echelle order
as CO bands). The $\chi^{2}$ value of LP\,412-31 is again very large
and this is due to unknown disturbance \footnote{Despite such a defect
in a part of the observed spectrum, modest number of the H$_2$O blends to be  
used in the mini CG analysis can be measured (table 7) outside the region 
shown in figure 11g.} on the observed spectrum as seen on figure 11g 
(hatched region).

\vspace{3mm}
\subsubsection{H$_2$O in  region A}
Based on log\,$A_{\rm C}^{(2)} $ and log\,$A_{\rm O}^{(3)}$ in tables\,6 and 
10\, respectively, H$_2$O spectra in  region A are computed with the
use of the specified models for our seven objects.
The resulting predicted spectra are compared with the observed ones
in figures\,12a-g. The $\chi^{2}$ values for the fitting are given in
the sixth column of table\,10.
The  $\chi^{2}$ values are large for GJ\,752B, GJ\,1002, and GAT\,1370,
and the $S/N$ ratios for these objects are again high (table 1,
fourth column).  The  $\chi^{2}$ value for LP\,412-31 is quite
small, and this may reflect  the very low $S/N$ ratio on one hand 
and good overall fit without major defect in this region
of the observed spectrum on the other.

\vspace{2mm}

--------------------

figure 10: p.19

figure 11: p.20

figure 12: p.21

-------------------- 

\section{Rotational broadening}
In the later M dwarfs, it is known that the fraction of the rapid rotators 
tends to be larger (e.g., \cite{Moh03}; \cite{Jen09}). Our object 
2MASS\,1835+32 
is also found to be  a fast rotator: The projected rotational velocity of
2MASS\,1835+32 was measured to be 44.0 $\pm$ 4.0\,km\,s$^{-1}$ \citep{Rei10}
and 37.6 $\pm$ 5.0\,km\,s$^{-1}$ \citep{Des12}. We adopt the mean value of
40.8\,km\,s$^{-1}$ in the following analysis, and examine the effect of
the uncertainty in the projected rotational velocity in section 7.4.
So far, we neglected the rotational broadening in our analysis of M dwarfs,
since it can be negligible compared to the rather large instrumental
broadening of our medium resolution spectrograph (FWHM $\approx$ 
16\,km\,s$^{-1}$). In the case of 2MASS\,1835+32, the rotational broadening
is much larger than the FWHM of the slit function, and we neglect it
for simplicity in the following analysis.
  
We first examine the effect of rotation on the predicted spectrum in 
the bandhead region of CO 2-0 band. The molecular
data used are the same as in Table\,7 \footnote{to which we add a blend 
with ref. no.\,15.  The spectroscopic data for the new blend are given in 
Appendix 1. The blend no.\,15 was not used so far since this blend is mostly 
located at the edge of the 28\,th  order of our echelle spectra. We use 
this blend in 2MASS\,1835+32, since otherwise we have only three blends 
for our analysis and, fortunately, this blend is well observed in this
object.} of Paper I. 
The spectrum is first calculated for our dusty model Bc2280c526 assuming
{\it case c} abundance, at the  sampling interval of 0.02\,\AA\,(resolution 
of $R \approx 10^{+5}$), and the result is shown by a thin line in 
figure\,13a. This spectrum is convolved with the rotation profile $A(x)$ 
given by Uns\"old (1955):
 $$  A(x) ={ { {2 \over \pi}(1-x^2)^{1 \over 2} + 
 {\beta \over 2}(1 - x^2) } \over { (1 + {2 \over 3}\beta) } }  \eqno(3)  $$
with
  $$ x ={ {\Delta\,\lambda} \over b }    \eqno(4)  $$
and
  $$  b = {\lambda \over c}\,V_{\rm rot}\,{\rm sin}\,i  \eqno(5)   $$
where $V_{\rm rot}$ is the equatorial rotational velocity, and $i$ is the
angle between the rotation axis and the line of sight to the observer.
Also, $\beta$ is the limb-darkening coefficient defined by
   $$  I = {\rm const} (1 + \beta {\rm cos}\theta)  \eqno(6) $$
where $I$ is the intensity emerging at an angle $\theta$  with respect
to the outward normal (we assume $\beta =2/3$) . The result convolved with the 
rotation profile of $ V_{\rm rot}\,{\rm sin}\,i =$ 40.8\,km\,s$^{-1}$ is 
shown by a thick line in figure\,13a.

For comparison, the same high resolution spectrum is convolved with the slit 
function (Gaussian) of FWHM = 16\,km\,s$^{-1}$, and the result is shown in 
figure\,13b. Comparison of figures\,13a and 13b reveals that most CO blends
with ref. nos. through 1 to 10 are smeared-out by the rotation and identities
of these blends are lost. Only blends with ref. nos. 11 -- 15 
conserve their identities and we will use the blends of ref. nos 12 - 15 in our
analysis of CO. The blend of ref. no. 11 is not used because the observed 
profile does not meet our criterion to accept a blend for our analysis (at 
least one wing reaches higher than the half of the line depth). 
It is to be noted that the pseudo-continuum level of
the rotationally broadened spectrum (figure 13a) is lower than that of the
broadened  by the slit function of the spectrograph (figure 13b). In
figure 13c, the observed spectrum of 2MASS\,1835+32 (dots) is compared 
with the rotationally broadened spectrum from figure 13a (thick line) after 
renormalized by the pseudo-continuum.

Next, we examine the effect of rotation on the H$_2$O spectrum. For this
purpose, we select the region between 22510 and 22660\,\AA, a part of
region B noted in Paper II (see its table 4). The high resolution spectrum 
is evaluated with the use of the line-list by BT2-HIGHTEMP2010 
(\cite{Bar06}; \cite{Rot10}) and the result is shown by a thin line 
in figure\,14a. The result convolved with the rotational profile of 
$ V_{\rm rot}{\rm sin}\,i =$ 40.8\,km\,s$^{-1}$
is shown by a thick line. For comparison, the same high resolution
spectrum convolved with the slit function of FWHM = 16\,km\,s$^{-1}$
is shown in figure\,14b on which the reference numbers introduced in 
Paper II are reproduced. Comparison of figures\,14a and 14b reveals 
that the identities of some H$_2$O blends are conserved despite the 
rotational broadening and some blends appear at slightly shifted
positions, including different components. We select eight features near
B01, 02, 03, 04 07, 09, 12, and 14 for our mini CG analysis as indicated 
in figure 14c, where the observed spectrum of 2MASS\,1835+32 (dots) is 
compared with the rotationally broadened spectrum of figure 14a (thick line).

But we could not
analyze the H$_2$O blends in region A noted in Paper II (see its  table 2),
since it is difficult to define the pseudo-continuum level for highly
depressed strong H$_2$O bands which are further smeared-out by rotation.   

\vspace{2mm}

--------------------

figure 13: p.22

figure 14: p.23

--------------------

\section{The M8.5 dwarf 2MASSI J1835379+325954}
In the M\,8.5 dwarf 2MASS 1832+35 in our sample, dust may form in its
photosphere and this object is also known to be a rapid rotator.
The analysis of the  CO and H$_2$O spectra is done by the mini
curve-of-growth method (subsection 7.1) and by the spectral synthesis
method (subsection 7.2) as for the seven earlier M dwarfs, but 
we apply the dusty models (section 4.2) and consider the rotational 
broadening (section 6). The effects of dust (subsection 7.3), 
rotational velocity (subsection 7.4), and  fundamental
parameters (subsection 7.5) are examined. Based on the examinations of
the various factors affecting the spectrum of 2MASS\,1835+32,
a best possible solution on the abundances and projected rotational
velocity in this dusty rapid rotator is suggested (subsection 7.6).

\subsection{Mini curve-of-growth analysis} 
 We apply the fully dusty model of $T_{\rm eff} = 2280$\,K, Bc2280c526,
assuming $ V_{\rm rot}{\rm sin}\,i =$ 40.8\,km\,s$^{-1}$ in our
analysis of the CO and H$_2$O blends in this subsection throughout.

\vspace{3mm}
\subsubsection{First iteration on the CO blends}
 We measure the equivalent widths of the CO blends with ref. nos. 12 - 15 
on the observed spectrum by referring to the pseudo-continuum and the 
results are given in the ninth column of table\,5. We again apply 
the mini CG method with the solar abundances
of {\it case c} as initial starting values (log\,$ A_{\rm C}^{(0)} =
-3.61$ and log\,$A_{\rm O}^{(0)} = -3.31 $), and determine the abundance
corrections needed to explain the observed log\,$W/\lambda$ values 
in table\,5. The resulting abundance corrections for the four lines are shown 
in figure 15a and the mean value  $\Delta$\,log\,$A_{\rm C}^{(1)} = -0.101$ 
is indicated by a dashed line. The resulting
carbon abundance is log\,$A_{\rm C}^{(1)} = -3.711 \pm 0.128$ 
(table\,11, line no.\,1).

\vspace{3mm}
\subsubsection{First iteration on the H$_2$O blends}
 We measure the equivalent widths of the H$_2$O blends near
B01, 02, 03, 04 07, 09, 12, and 14 (see figure 14c) on the observed spectrum 
by referring to the pseudo-continuum and the results are given in the
ninth column of table\,7. We again apply the mini CG method to these blends. 
Since carbon abundance is now known to be log\,$A_{\rm C}^{(1)} = -3.711$, we
estimate log\,$A_{\rm O}$ to be larger by 0.30 dex as in the Sun 
\citep{All02}. Thus we start from log\,$A_{\rm O}^{(0)} =-3.711 +0.30 =
-3.411 $ and since this value may be a good starting value,  we now apply the 
mini CG method with $\delta = 0.1$. The resulting abundance corrections
from the eight blends are shown in figure\,16a and the mean value 
$\Delta$\,log\,$A_{\rm O}^{(1)} = +0.008$ is indicated by
a dashed line. The result is log\,$A_{\rm O}^{(1)} = -3.403 \pm 0.024$ as 
given in table\,11 (line no.\,2). 

\vspace{3mm}
\subsubsection{Confirmation by second iteration}
Given that we have now determined log\,$A_{\rm C}^{(1)}$ and 
log\,$A_{\rm O}^{(1)}$, we repeat the mini CG analysis with these values as 
starting values. Since the CO blends are contaminated with the weak H$_2$O 
lines, revised oxygen abundance may have some effect on the determination of
carbon abundance from the CO blends. The resulting mean abundance corrections
is only $-0.025$ (table 11, line no.\,3), confirming 
that the first iteration is already close to the solution, and this
may be because the oxygen abundance assumed in the first iteration
is already close to the final result (as will be confirmed in table\,11). 
 With the revised carbon abundance, the H$_2$O blends are  also examined
and the resulting  mean correction is only $-0.012$ (table 11, line no.\,4). 

We confirm that the first iteration is already near the convergence value.
For this reason,  we will skip the second iteration in the analysis to be
given hereafter (subsections 7.3, 7.4, and 7.5), since these analyses are for 
comparison purpose to examine the effect of different assumptions. For 
consistency in these analyses,  we will use the result of the first iteration 
given in table 11 (line nos.\,1 and 2) as the reference. 

\vspace{2mm}

--------------------

figure 15: p.24

figure 16: p.24

\vspace{2mm}

table 11: p.33

--------------------

\subsection{Synthetic spectra}
Based on the  carbon and oxygen abundances given in table\,11 (line nos.\,1 
and 2), synthetic spectra of CO and H$_2$O are computed and compared with
the observed spectra in figures 17a  and 18a, respectively. Although only
four CO blends are used to obtain the carbon abundance from the CO
spectrum, the whole CO spectrum including the bandhead region
is reasonably well reproduced with the abundances based on the
mini CG analysis. The $\chi^2$ values for CO and H$_2$O are 5.803 and 2.035
(table 12, line no.\,1), respectively \footnote{Although we notice 
in section 5.2 that the  $\chi^2$ values are not so useful as measures of
the goodness of fits if applied to different objects having different $S/N$ 
ratios, they are  useful if applied to the same object analyzed 
by different models.}.

\vspace{2mm}

--------------------

figure 17: p.25

figure 18: p.26

\vspace{2mm}

table 12: p.33

--------------------

\subsection{Effect of dust on abundance determination}
 We apply our dusty model in our abundance analysis in  subsections\,7.1 and
7.2. We now examine the case of a dust-free model to assess the effect of dust
in our analysis. For this purpose, we use the  dust-free model of the same
values of $T_{\rm eff}$ = 2280\,K, log\,$g$ = 5.26, and 
$ V_{\rm rot}\,{\rm sin}\,i = $ 40.8\,km\,s$^{-1}$ as the dusty model we
applied in  subsection\,7.1. We assume the same initial abundances of 
log\,$A_{\rm C}^{(0)} = -3.61$ and log\,$A_{\rm O}^{(0)} = -3.31$, and 
apply the mini CG method to the CO blends as for the dusty model. The 
resulting abundance corrections are shown in figure\,15b. The mean value
of $\Delta$\,log\,$A_{\rm C}^{(1)} = -0.358 \pm 0.078$ is 0.257\,dex 
smaller compared with the result of $\Delta$\,log\,$A_{\rm C}^{(1)} = -0.101
\pm 0.128$ using the dusty model. 

Then, we assume the 
revised carbon abundance log\,$A_{\rm C}^{(1)} = -3.61 -0.358 = -3.968$ and
log\,$A_{\rm O}^{(0)} = -3.968 +0.3 = -3.668$ as starting values, and
analyze the H$_2$O blends. The resulting abundance corrections are shown 
in figure\,16b and the mean value of  $\Delta$\,log\,$A_{\rm O}^{(1)} = 0.144
\pm 0.034$ is compared with  the result using the dusty model, 
$\Delta$\,log\,$A_{\rm O}^{(1)} = +0.008 \pm 0.024$. However, the starting 
values are different from those of
the dusty case this time, and what is to be compared is the resulting
log\,$A_{\rm O}^{(1)} = -3.668 +0.144 = -3.524$ of the dust-free case with   
log\,$A_{\rm O}^{(1)} = -3.411 + 0.008 = -3.403$ of the dusty case. 
Summarizing, the carbon and oxygen abundances based on the dust-free model
are log\,$A_{\rm C}^{(1)} = -3.968 \pm 0.078 $ and log\,$A_{\rm O}^{(1)} 
= -3.524 \pm 0.034$ (table 11, line nos.\,5 and 6), or smaller by 0.257 
and 0.121\,dex, respectively, compared with the case by the dusty model 
(table 11, line nos.\,1 and 2). 

We also compute the synthetic spectra of CO and H$_2$O,  and the 
results are shown in figures\,17b and 18b,
respectively. We also evaluate the $\chi^2$ values which turn out to
be $\chi^2$(CO) = 10.369 and $\chi^2$(H$_2$O) = 2.879 by the dust-free
model (table\,12, line no.\,2), compared with $\chi^2$(CO) = 5.803 and 
$\chi^2$(H$_2$O) = 2.035  by the  dusty model (table\,12, line no.\,1).  
Thus, the abundances based on the dusty model
provide a better account of the observed spectra both of CO and H$_2$O.
Also, the resulting carbon and oxygen abundances based on the dust-free
model appear to be somewhat too small for an M dwarf of the disk
population. For these reasons, we may conclude
that the  dusty model better represents the photospheric structure of
2MASS 1832+35 and that it is more appropriate for abundance analysis.

\subsection{Effect of the rotational velocity}
So far, we apply the projected rotational velocity of $V_{\rm rot}{\rm sin}\,i$
= 40.8\,km\,s$^{-1}$, which is a mean value of the two determinations by
different groups (section\,6).  During our analysis, we notice that
the projected rotational velocity has appreciable effect on the abundance
determination. This is because the EWs are evaluated from the synthetic 
spectrum by referring to the pseudo-continuum level which  depends on 
the rotational broadening, in our mini CG analysis. To examine the effect
of the rotational velocity,
we apply the mini CG method to the CO blends by assuming the lower value of 
$ V_{\rm rot}{\rm sin}\,i =$ 37.6\,km\,s$^{-1}$ \citep{Des12} and the higher 
value of $ V_{\rm rot}{\rm sin}\,i =$ 44.0\,km\,s$^{-1}$ \citep{Rei10},
both using the dusty model Bc2280c526 with the initial starting abundances of
log\,$A_{\rm C}^{(0)} = -3.61$ and log\,$A_{\rm O}^{(0)} = -3.31$. The
resulting abundance corrections are shown in figure\,15c and 15d for
the cases of the low and high rotational velocities, respectively.
The resulting carbon abundances are log\,$A_{\rm C}^{(1)} = -3.794
\pm 0.094$ and $-3.566 \pm 0.199$ (table 12, line nos.\,7 and 9) for the 
low and high rotational velocities, respectively. Thus the difference of 
the resulting carbon abundances is 0.228\,dex for the difference of   
$ V_{\rm rot}\,{\rm sin}\,i $ values by 6.4\,km\,s$^{-1}$.

We also repeat the mini CG method for the H$_2$O blends and the results
are shown in figures\,16c and 16d for the cases of the low and high rotational 
velocities, respectively. The resulting oxygen abundances are
log\,$A_{\rm O}^{(1)} = -3.461 \pm 0.036$ and $ -3.318 \pm 0.029$
(table 12, line nos.\,8 and 10) for the cases of the low and high 
rotational velocities, respectively, and 
the difference is 0.143\,dex for the difference of   
$ V_{\rm rot}{\rm sin}\,i $ values by 6.4\,km\,s$^{-1}$.

The synthetic spectra of CO are shown in figures\,17c and 17d for
the cases of the low and high rotational velocities, respectively.
The observed spectrum remains unchanged in figures\,17c and 17d, but
the pseudo-continuum level of the synthetic spectrum is  
lower \footnote{The spectra in figures\,17c and 
17d are normalized already by the pseudo-continua. We confirm that
the pseudo-continuum levels are 0.929  and 0.917  of the true continua 
in the low and high rotational velocities, respectively.}
in the case of the higher rotational 
velocity than in the case of the lower rotational velocity for the reason
noted at the beginning of this subsection. For this reason, EWs of the 
CO blends measured from the synthetic spectrum of the higher rotational
velocity are smaller than those measured from the spectrum of the lower
rotational velocity, as can directly be seen on figures\,17c and 17d. 
Then, a higher abundance is required to explain the observed spectrum by 
the predicted spectrum based on the higher rotational velocity. 
The fitting appears to be  better for the case of the higher rotational 
velocity for which the $\chi^2$ value is 3.126 while it is 9.406 for the case 
of the lower rotational velocity (table 12, line nos.\,3 ad 4).
 
The similar analyses on the H$_2$O blends are shown in figures\,16c and 16d
for the  cases of the low and high rotational velocities, respectively.
The fittings  appear to be somewhat better for the case of the higher 
rotational velocity, since the $\chi^2$ values are 2.568 and 1.820 for the 
low and high rotational velocities, respectively (table\,12, line nos.\,3 
and 4). By the way, $\chi^2$(CO) = 5.803 and $\chi^2$(H$_2$O) = 2.035 
for the case of $ V_{\rm rot}{\rm sin}\,i $= 40.8\,km\,s$^{-1}$ discussed in
subsection 7.2. In conclusion, the highest rotational velocity of 
 $ V_{\rm rot}{\rm sin}\,i $= 44.0\,km\,s$^{-1}$ provides the best account
of the observed spectra (compare line no.\,1, 3, and 4 of table 12), and 
possibly the best carbon and oxygen abundances.

\subsection{Effect of the fundamental parameters}
Finally, we examine the effect of the uncertainty in the fundamental
parameters, especially of $T_{\rm eff}$ by $\pm 100$\,K. For this purpose,
we generate new dusty models of $T_{\rm eff}$ = 2180 and 2380\,K. The gravities
are estimated from the $T_{\rm eff}$ - radius - mass relations discussed
in section\,3. We apply our dusty models of the low and high values of 
$T_{\rm eff}$, Bc2180cc528 and Bc2380c525, respectively, with the initial 
starting abundances of log\,$A_{\rm C}^{(0)} = -3.61$ and 
log\,$A_{\rm O}^{(0)} = -3.31$. We also assume $ V_{\rm rot}{\rm sin}\,i 
=$ 40.8\,km\,s$^{-1}$. The resulting carbon abundance corrections by the 
mini CG  analysis for the low and high $T_{\rm eff}$ models  are
shown in figures\,15e and 15f, respectively. It appears that the resulting
carbon abundance corrections are $\Delta$\,log\,$A_{\rm C}^{(1)} = +0.0003$ 
$\pm 0.136 $ and $-0.094 \pm 0.142$ (table 11, line nos.\,11 and 13)
 for the low and high $T_{\rm eff}$, respectively. For comparison, 
$\Delta$\,log\,$A_{\rm C}^{(1)} = -0.101$ $\pm 0.128$ for our adopted case
of  $T_{\rm eff}$ = 2280\,K (table 11, line no.\,1).
Thus the changes of $T_{\rm eff}$  by $\pm\,100$\,K result in rather
minor changes on the resulting carbon abundances. This should certainly
be due to special circumstance noted in section\,4.3 that the temperatures
of the surface layers converge to the condensation lines of iron.

We repeat the same analysis on H$_2$O blends and the resulting
abundance corrections are shown in figures\,16e and 16f for
the cases of the low and high $T_{\rm eff}$, respectively.
The resulting oxygen abundances are log\,$A_{\rm O}^{(1)} = -3.324 \pm 0.023$
and $-3.388 \pm 0.023$ (table 11, line nos.\,12 and 14) for  
$T_{\rm eff}$ = 2180 and 2380\,K, respectively. For comparison, 
log\,$A_{\rm O}^{(1)} = -3.403 \pm 0.024$ for  $T_{\rm eff}$ = 2280\,K 
(table 11, line no.\,2). Again, the oxygen abundance depends little on 
$T_{\rm eff}$. This is a favorable result, since the effective temperature is 
difficult to determine in very cool M dwarfs since no angular diameter
measurement is available yet. 

We then compute the synthetic spectra of CO for the abundances determined
for the  cases of the low and high $T_{\rm eff}$  and the results are shown
in figures\,17e and 17f, respectively. 
 We also compute the synthetic spectra of H$_2$O for the abundances determined
for the  cases of the low and high $T_{\rm eff}$  and the results are shown
in figures\,18e and 18f, respectively.  The results are again not much 
different compared with the reference case,  
and the $\chi^2$ values are nearly the same for three cases (compare line 
nos. 1, 5, and 6 of table 12).

\subsection{Carbon and oxygen abundances in 2MASS\,1835+32}
  In the latest M dwarf 2MASS\,1835+32 in our sample, the dust formation and 
rapid rotation introduce additional uncertainty in the abundance 
determinations. The effect of the dust formation is  appreciable
(subsection 7.3), but this is based on the simplest assumption on
the dust formation based on the thermodynamics alone. The limitation of 
such a simple approach will be discussed in section 8.2, but 
we think that even our simple treatment of  dust  provides a 
better account of the observed spectrum of 2MASS\,1835+32 compared
with the case based on the dust-free model, and we decide 
to apply our dusty models in our abundance analysis of this object.

  The effect of the rotational velocity turns out to be unexpectedly large. 
We first apply the mean value ($ V_{\rm rot}{\rm sin}\,i $= 40.8\,km\,s$^{-1}$)
of the two available values in the literature (section 6), but we find that
the higher value ($ V_{\rm rot}{\rm sin}\,i $ = 44.0\,km\,s$^{-1}$) of the 
two literature values gives the better account  of the observed spectrum 
(subsection 7.4). The empirical determinations of the rotational velocity
are anyhow difficult in late M dwarfs since the true-continuum level of
the observed spectrum cannot be well defined in general and, further, 
abundances and physical parameters
(or model photosphere applied) must be properly assigned. We believe that
our analysis can be a way to clear these requirements. Since our analysis
suggests a higher value of the known $ V_{\rm rot}{\rm sin}\,i $ values, 
we examine a still higher value of   
$ V_{\rm rot}{\rm sin}\,i $ = 48.0\,km\,s$^{-1}$ and repeat the
analysis outlined in subsection 7.4. The results are log\,$ A_{\rm C} = 
-3.451 \pm 0.298 $ and  log\,$A_{\rm O} = -3.245 \pm 0.044$. However,
we can use only four lines of CO, and we notice that the internal consistency
(p.e.) of our CO analysis turns out to be worse for the higher 
$ V_{\rm rot}{\rm sin}\,i $ value. Also we find that $\chi^2$ = 3.476 for 
the comparison of the observed and predicted spectra of CO, and 
$ V_{\rm rot}{\rm sin}\,i $ = 48.0\,km\,s$^{-1}$ does not provide any 
improvement over the result based on 44.0\,km\,s$^{-1}$ for which  
$\chi^2$ = 3.126 (table 12, lone no.\,4). 

The abundances of log\,$A_{\rm C} = -3.711$ and  log\,$A_{\rm O} = -3.403$ 
based on $ V_{\rm rot}{\rm sin}\,i $ = 40.8\,km\,s$^{-1}$ (table 11, line 
nos.\,1 and 2) are somewhat too low for a possibly young object like 
2MASS\,1835+32 which is not spun down yet. On the other hand, the 
abundances of log\,$A_{\rm C} = -3.566 $ and  log\,$A_{\rm O} = -3.318$ based
on $ V_{\rm rot}{\rm sin}\,i $ = 44.0\,km\,s$^{-1}$ (table 11, line nos.\,9 
and 10) appear to be  acceptable for 2MASS\,1835+32 which may be 
a relatively young M dwarf as noted above. 
Also, inspection of table 12 reveals that this case provides  
the best account of the observed spectrum among the six possible cases we have 
examined. For these reasons, we adopt log\,$A_{\rm C} = -3.57 \pm 0.20 $ and  
log\,$A_{\rm O} = -3.32 \pm 0.03$ together with $ V_{\rm rot}{\rm sin}\,i $
= 44.0\,km\,s$^{-1}$ for 2MASS\,1835+32 (table 13).

\section{Discussion}
\vspace{3mm}

\subsection{Carbon and oxygen abundances in M dwarfs}
  We take the weighted mean (with the numbers of lines used as weights)
of the oxygen abundances resulting from the
analyses of regions A (table 10) and B (table 8) for the seven M dwarfs
for which the two regions have been analyzed. The results and
that from  region B alone (2MASS 1835+32) are summarized in table\,13
together with the carbon abundances from tables\,6 and 11 (line nos.\,9 and
10). Also, the classical and more recent solar carbon and oxygen abundances 
by \citet{And89} and \citet{Asp09}, respectively, are included in table\,13 
for comparison.

The resulting values of log\,$A_{\rm O}/A_{\rm C}$ (given in the fourth
column of table\,13) are plotted against the values of
log\,$A_{\rm C}$ in figure\,19 by red (or black) filled circles. 
The  results for 38 M dwarfs analyzed in Papers I \& II are reproduced 
from figure\,15 of Paper II, as shown by green (or grey)  filled circles.
The two representative solar values are  shown by $\solar$ marks. 
Inspection of figure 19 reveals that the late M dwarfs show no systematic 
difference from the early and middle M dwarfs, and  our conclusion 
outlined in Paper II that the $A_{\rm O}/A_{\rm C}$ ratios are larger at 
the lower metallicities and gradually decrease in the higher metallicities is 
strengthened with the additional data on the eight late M dwarfs. 

Although the large production of oxygen in the
metal-poor era is explained as due to Type II supernovae/hypernovae in the 
early Galaxy \citep{Nom13}, the large differential effect in the
productions of carbon and oxygen is by no means well understood \citep{Gus99},
and  accurate determinations of the carbon and oxygen abundances will still 
be needed. Especially, our analysis is limited to the M dwarfs in the
disk, and to extend our analysis to the halo M dwarfs (M subdwarfs)
is a next major step. However, besides the observational difficulty to 
observe faint M subdwarfs, the $K$ band region of the metal deficient M dwarfs
is badly depressed by the strong collision-induced absorption (CIA) due to
H$_2$ - H$_2$ and H$_2$ - He pairs \citep{Bor97}. In fact, the CO and
H$_2$O bands we have analyzed in our present work are almost unseen on the
observed spectra of a few late M subdwarfs, at least by low 
resolutions \citep{Bur06}. For this reason, abundance analysis of
M subdwarfs will be quite challenging even if  higher resolution spectra of 
M subdwarfs can be obtained by future large telescopes.

By the way, the effect of H$_2$ CIA on the abundance analysis is by no 
means well evaluated so far as we are aware, and we examine this problem 
in the case of a relatively cool M dwarf GJ\,752B as an example. 
In figure\,20, we plot the predicted
spectrum of CO based on our model of GJ\,752B (Cc2640c526) but disregarding
the H$_2$  CIA due to H$_2$ - H$_2$ and H$_2$ - He pairs (blue/black solid 
line), and compare it with that based on the same model but including 
the H$_2$ CIA (light sky/grey solid line) as we have
done so far (i.e., the same as figure\,10f). The resulting CO spectrum
computed disregarding the H$_2$ CIA is stronger than that including the
H$_2$  CIA, as expected. Then, we apply the mini CG analysis based on 
our model of GJ\,752B but disregarding the H$_2$ CIA to the CO lines,
and determine the abundance correction to the log\,$A_{\rm C}$ based on
the model including the H$_2$  CIA (log\,$A_{\rm C} = -3.550 \pm 0.066$,
see Table 6). The result is $\Delta {\rm log}\,A_{\rm C} = -0.196 \pm 0.081$
as shown in figure\,21, and the effect of neglecting the H$_2$  CIA is
to reduce the logarithmic carbon abundance by 0.196 dex. Thus, the effect
of the H$_2$ CIA is appreciable even in the solar metallicity case, even if
it is not so drastic as in the metal-poor cases \citep{Bor97}.

\vspace{2mm}

--------------------

figure 19: p.27

figure 20: p.27

figure 21: p.28

\vspace{2mm}

table 13: p.34

--------------------

\subsection{Dust in the photospheres of M dwarfs}

In section 4.2, we assume  that dust forms everywhere in the photospheres 
of M dwarfs so long as the thermochemical condition for condensation is 
fulfilled. However, it is now  known that dust does not fill in the
whole photosphere where thermochemical condition of
condensation is fulfilled, at least in L and T dwarfs. In fact, if  dust 
forms this way, more dust should be formed in the cooler dwarfs, and hence 
the cooler dwarfs should increasingly be red. However, observations never 
follow such an expectation. For example, $J-K$ color first shows reddening 
from late M to L dwarfs, but it turns to blue after late L  dwarfs
(e.g., \cite{Kna04}), suggesting that dust in the photospheres 
should decrease towards late T dwarfs.

We have developed a simple model referred to as the unified cloudy model, 
UCM \citep{Tsu02}: Dust grains first 
form at $T_{\rm cond}$, the condensation temperature,
but dust grains grow larger at cooler temperature and will
precipitate at a certain temperature which we referred to as
the critical temperature, $T_{\rm cr}$. Then, dust grains
exist only in the limited region of $ T_{\rm cr} \lesssim T \lesssim
T_{\rm cond} $ or dust grains appear in a form of cloud.
In an L dwarf, $ T_{\rm eff} $ (approximately the temperature where
optical depth is about unity) is sufficiently higher than
$T_{\rm cond}$ (see figure 5) as well as $T_{\rm cr}$
(1700, 1800, 1900\,K and $ T_{\rm cond}$). For this reason, the dust cloud
appears in the optically thin region and the L dwarf appears dusty.
On the other hand, $T_{\rm eff}$ of a T dwarf is generally lower than
$T_{\rm cond}$ as well as $T_{\rm cr}$, and the dust cloud is
situated in the optically thick region. For this reason, the dust cloud
will have little observable effect and the T dwarf appears 
observationally as if it is dust-free. 
Then,  observed characteristics of the spectra of L and T dwarfs could 
naturally be accounted for  as a single sequence of $T_{\rm eff}$ in 
our UCMs. 

However, further progress in photometry of L and T dwarfs 
revealed that the infrared colors such as $ J-K $ \citep{Kna04} show 
a large variation at a given effective temperature \citep{Vrb04}.
This fact suggests that the infrared colors are not determined by
the  $T_{\rm eff}$  alone, and we proposed that the
critical temperature, $T_{\rm cr}$, introduced in our UCMs, should
be changing at a given $T_{\rm eff}$ \citep{Tsu05}. Since the value of 
$T_{\rm cr}$ essentially defines the thickness of the dust 
cloud \footnote{In our UCM, dust cloud is located in the regime of
 $T_{\rm cr} \lesssim T \lesssim T_{\rm cond}$, where $T_{\rm cond}$ is 
uniquely determined by thermodynamics while $T_{\rm cr}$ is a free 
parameter to be fixed empirically. Then, thickness of the dust cloud  is 
larger if the difference of  $T_{\rm cr}$ against $ T_{\rm cond}$ is larger.},
we concluded that the thickness of the dust cloud should be changing at
a fixed $T_{\rm eff}$. 
Further, the change of the spectra  from L6.5 to T3.5, 
for example, could be explained as a result of increasing  $T_{\rm cr}$
from 1700\,K to $T_{\rm cond}$ 
while $ T_{\rm eff} $ remains nearly constant 
at $T_{\rm eff} \approx 1300$\,K (see Fig.10 of \cite{Tsu05}).
This fact implies that the spectra as well as infrared colors of 
L and T dwarfs are not determined primarily by $ T_{\rm eff} $ but should
be determined by $ T_{\rm cr} $ or the thickness of the dust cloud.
Given that dust has such  large and definitive effect on the observed spectra 
and colors, the values of $T_{\rm cr}$ can be inferred  from  
observations in L and T dwarfs. 

Although our cloudy models could be  applied to L and T dwarfs, 
we are not sure if the same cloudy models can be applied to
late M dwarfs. In fact, we have no means by which to estimate the value of  
$T_{\rm cr}$ in the case of M dwarfs, since the effect of dust on the
observed spectra is rather subtle compared with that in L and T dwarfs. 
For this reason, we do not apply the cloudy models included in our UCMs to M
dwarfs, but apply the simple fully dusty models of case B, also included as a
subset in our UCM grid. If dust grains form a thin cloud in M dwarfs, our
treatment provides a maximun estimate of the effect of dust.

Such a limitaion of our simple approach basically based on the
thermodynamics alone has been overcome by the recent
development of the self-consistent treatment of dust formation  
coupled with the photospheric structure (\cite{Hel08b}; \yearcite{Hel08c}). 
These authors followed the kinematic processes of nucleation, 
dust growth, and evaporation, resulting in  gravitational 
setting in the atmosphere,
and the resulting dust cloud structure was incorporated into the
photospheric model structure. They have applied their method to
the substellar atmospheres for a wide range of metallicity extending
to [M/H] as low as -6.0 and showed that dust clouds form even in the
most metal-poor substellar objects \citep{Wit09}. 
They also have applied their method to evaluate
the synthetic spectra of cool dusty dwarfs including late M dwarfs, and 
showed that the observed infrared spectra of late M and L dwarfs  could be
well fitted with their model predictions (Witte et al., 2011). 
Such a success is quite encouraging for the spectral analysis of
dusty dwarfs, and we hope that their method will be extended to the 
abundance analysis on high resolution spectra of the dusty  M  dwarfs and
subdwarfs in the near future. 

One important effect of dust formation on abundance analysis is 
the element depletion by dust grains.
We  examine this problem within the framework of our 
 LTE model and estimate the effect of dust formation on 
the abundance determination
in our case of 2MASS 1825+32 based on the photosheric model Bc2280c526.
For this purpose, we plot in figure\,22 the logarithm of the fraction 
$f_{\rm mol}$ of oxygen atoms in the species (including molecules and 
dust grains formed in the photosphere) defined by
       $$ f_{\rm mol} = q\,P_{\rm mol}/P({\rm O}), \eqno(7) $$
where $P_{\rm mol}$ \footnote{For a solid particle such as corundum, 
for example, $P_{\rm Al_2O_3}$ is the
fictitious pressure of the monomer Al$_2$O$_3$ that would appear when 
 corundum is fully dissolved into its constituent monomers.} 
is the partial pressure of a molecule and $q$ is 
the number of oxygen atoms in the chemical formula of the molecule.
Also,  $P({\rm O})$  is the fictitious pressure of oxygen nuclei 
        $$ P({\rm O}) = A_{\rm O} P({\rm H}) , \eqno(8)   $$
with   the fictitious pressure of hydrogen nuclei $P({\rm H})$ obtained from
        $$ P({\rm H}) = P_{\rm H^{+}} + P_{\rm H} + 2 P_{\rm H_2}, \eqno(9)  $$
where $P_{\rm H^{+}}$, $P_{\rm H}$, and $P_{\rm H_2}$ are the partial 
pressures of H$^{+}$, H,  and H$_{2}$, respectively, in the photosphere. 

Inspection of figure\,22 reveals that oxygen atoms are mostly
depleted by CO, H$_2$O, and SiO.  We consider only three dust species 
(corundum, iron, and enstatite) , and the dust grain 
that works as a sink of oxygen is only corundum (Al$_2$O$_3$), since
enstatite (MgSiO$_3$) does not form in our dusty model yet (see figure\,5). 
In the abundance analysis based on H$_2$O, the fraction of oxygen
that is depleted by corundum  must be corrected for, but
we skipped this correction since the
corundum consumes only about 1\% of oxygen and hence its effect
on the oxygen abundance is about 1\%. If enstatite forms, its effect
on the oxygen abundance should be larger, since Si and Mg consume more
than  10\% of oxygen because of their higher abundance (about 10\%
of the oxygen abundance). For this reason, we are considering enstatite 
as a sink of oxygen in our chemical equilibrium, but this effect is not
shown in figure\,22, since enstatite does not
 form in our model of the late M dwarf 2MASS 1835+32 (see figure 5). Then,
the uncertainty in our oxygen abundance due to dust formation may be
about 1\% within the framework of our present LTE analysis. 
Certainly, the dust species considered by us are 
limited (only three) and the effect of element depletion should be
different if more dust species are considered \citep{Hel08c}.
But the amount of the oxygen depletion, for example,
is limited by the abundances of non-volatile elements (Al, Mg, Si, etc.)
that combine with oxygen and we hope that an approximate effect
of dust formation on abundance analysis can be represented by
the species that consumes the larger amount of oxygen atoms (e.g. corundum
in our case discussed above under the absence of enstatite).
 However, it is certainly necessary to consider more dust species if
higher accuracy in abundance determination is required.

\vspace{2mm}

--------------------

figure 22: p.28

--------------------

\subsection{The end of the main sequence}
    Probably, our present sample includes some objects near the end 
of the main sequence. The latest ones, such as GAT\,1370, LP\,412-31,
and 2MASS 1835+32 are even referred to as brown dwarfs in SIMBAD,
but the reason for these assignments is by no means clear for us.
Unfortunately, we do not know how to exactly discriminate  a brown dwarf 
from  a star for  a particular object in question.   
This is because there is no definite spectral criterion by which to
identify the substellar nature. The Li test \citep{Rob92} applied to
field objects is effective in confirming the brown dwarf status of objects
less massive than 0.06\,$M_{\odot}$, while the three objects mentioned above
are likely to be more massive than 0.06\,$M_{\odot}$ (table 3). This method
requires reasonably high spectral resolution and may be difficult to be 
used as a spectroscopic criterion in spectral classification level.   
Spectral classification of late M dwarfs, however, is also hampered
by  the presence of dust in their photospheres. Unlike
atomic and molecular lines that show definitive signatures of their
origins, dust spectra have no such signature.  For this reason,
it is not yet known  how the effect of dust can be taken into account into
 the spectral classification of late M dwarfs. 
Anyhow, it is  difficult to define the end of the main-sequence
observationally and  we simply follow the spectral classification that 
has classified our objects as M dwarfs in the present work.

On the other hand, the observed HR diagram at the end of the 
main sequence is  well explained by the evolutionary model
by \citet{Bar98} (section 6.4 of Paper I) and consistent with their 
theoretical HR diagram (see figure 17 in Paper I).  Also, some physical
parameters predicted by their model are well consistent with
the observed values (figures 2 - 4). For this reason,
our estimations of some physical parameters are helped by their evolutionary 
model (section 3). Also,
chemical abundances are shown to be rather normal down to the end of 
the main sequence, even though our analysis has been limited to the carbon 
and oxygen abundances yet, and our result appears to be consistent with 
the present idea of the Galactic chemical evolution (e.g., figure 19).

\section{ Concluding remarks}

Our spectroscopic analysis of M dwarfs is greatly helped by the recent 
progress in observations, including the angular diameter measurements
by the interferometry (e.g., \cite{Boy12}), infrared spectroscopy using 
new detectors  (e.g., \cite{Kob00}), and high precision astrometry
(e.g., \cite{Lee07}) and photometry (e.g., \cite{Wri10}) from  space.
A remaining  problem in the present paper is that the effective
temperatures of the latest M dwarfs cannot be determined by 
directly measured angular diameters, and we hope that  the angular
diameter measurements can be extended to late M dwarfs in the near future.

We have determined the carbon and oxygen abundances in 38 + 8 = 46 M dwarfs
from the CO and H$_2$O spectra through our Papers I, II, and III, and the
carbon abundances in additional four early M dwarfs in Paper I. So far,
M dwarfs might not be expected to be proper objects for abundance 
determinations in general, but
we now believe that such a prejudice should be reconsidered. In fact,
we think that the stellar carbon and oxygen abundances can best be
determined from the numerous CO and H$_2$O lines, respectively,
observed in the spectra of M dwarfs, especially because stable CO and 
H$_2$O molecules are the major species of carbon and oxygen, respectively, in
M dwarfs and hence their abundances are almost unaffected by  
the uncertainty in  photospheric structures (Paper II). For this reason, 
the resulting carbon and oxygen abundances are insensitive to the
imperfect  model photospheres used in the abundance analysis.
This fact shows a marked contrast to the well known difficulties in
determining carbon and oxygen abundances from other molecular or atomic
lines usually pretty model-sensitive in solar-type stars. An example
of such an  advantage is that the atypical nature of the recently revised 
solar carbon and oxygen abundances \citep{Asp09} for its metallicity, 
compared with the nearby unevolved stars, has been 
 demonstrated clearly by our analysis (see figure\,15 of Paper I and 
figure\,16 of Paper II for C and O, respectively), 
because of the  higher internal consistency of our analysis based
on M dwarfs compared with the other works based on the solar-type stars.   

Also,  even though the true-continuum cannot be seen in the spectra of
M dwarfs, the pseudo-continuum defined by numerous lines of H$_2$O 
can be well defined observationally and, at the same time, it can 
theoretically be  evaluated fairly  accurately   thanks to the recent 
high precision line list of H$_2$O (e.g., \cite{Bar06}; \cite{Rot10}).
Then,  the basic principle of the quantitative analysis of the stellar spectra 
can be applied essentially in the same way without regard to whether 
the true- or the pseudo-continuum is referred to. This way, the difficulty 
of the continuum in cool stars is 
resolved at least in the near infrared spectra of M dwarfs (Paper I). 
Also, our analysis is not necessarily limited to the well defined 
unblended lines but can be extended to any blends by the use of a flexible  
method such as the mini curve-of-growth analysis.   

 In late M dwarfs, dust formation and rotation introduce additional problems
in the spectroscopic analysis. Especially, effect of dust on  abundance
analysis involves complicated problems such as cloud formation, element 
depletion by dust grains, dust properties and so on, and impressive progress 
is being achieved in this field (section\,8.2). In the present paper, 
however, we have restricted our analysis to a simple treatment based on the 
thermodynamical argument, and we hope that the effect of dust formation
on abundance determination can be estimated approximately. Except 
for the dusty M dwarfs, CO and H$_2$O remain as excellent 
abundance indicators of carbon and oxygen, respectively, even in
late M dwarfs (Paper III). 

What we have done in our present work, however, is limited  to the carbon 
and oxygen abundances determined from the CO and H$_2$O spectra, respectively,
and it is true that this is due  to some favorable conditions in these
particular cases. It is still to be investigated if similar analysis can 
be extended to other molecular or atomic spectra, but it is
by no means trivial if such an extension can be done easily. However, 
given the recent progress in observations and in basic data such as 
molecular data, we believe that stellar spectroscopy will have great
possibility  in uncovering abundant information coded in high resolution 
stellar spectra.

\bigskip

We thank an anonymous referee for careful reading of our text and
for helpful suggestions, especially on the cloud formation in 
dusty M dwarfs.

We thank  the staff of the Subaru Telescope and S. Sorahana for the
help with observations.

This research makes use of data products from the Wide-field Infrared
Survey Explorer which is a joint project of the University of
California, Los Angeles, and the Jet Propulsion 
Laboratory/California Institute of Technology, funded by NASA.

This research has made use of the VizieR catalog access tool and
the SIMBAD database, both operated at CDS, Strasbourg, France, and
of the RECONS database in www.recons.org.

Computations are carried out on common use data analysis
computer system at the Astronomy Data Center, ADC, of the National
Astronomical Observatory of Japan.

\appendix

\section{Additional spectroscopic  data of CO lines  (2-0 band)}
   In addition to the CO blends referred to as nos.\,1 - 14 in table 7 of
Paper I, we  use an additional blend referred to as no.\,15 in this
paper and the spectroscopic data for this blend composed of two  CO lines 
are given in table 14.

\vspace{2mm}

--------------------

table 14: p.34

--------------------

\clearpage

\begin{figure}
   \begin{center}
       \FigureFile(84.4mm,0mm){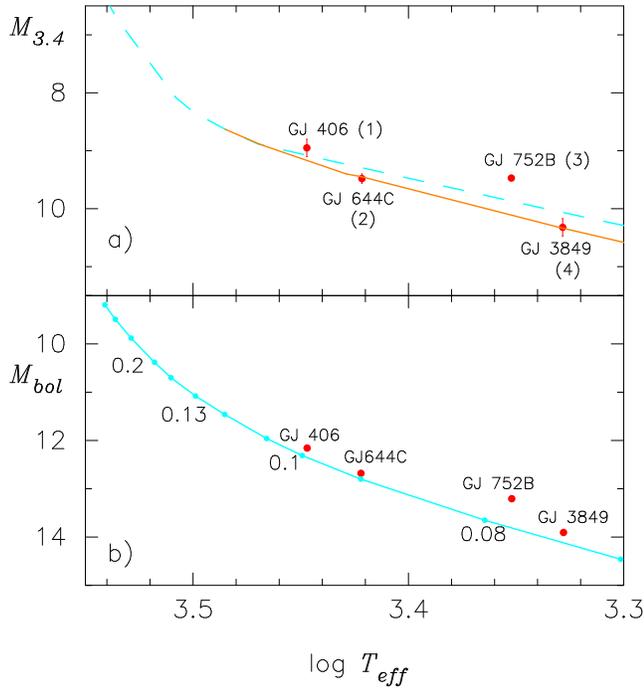}
   \end{center}
   \caption{ a) The absolute magnitude at 3.4\,$\mu$m, $M_{\rm 34}$,
plotted against log\,$T_{\rm eff}$ for four late M dwarfs, where
$T_{\rm eff}$ values are based on the infrared flux method
and a dashed line is a mean curve from figure 1 of Paper I.
b) The absolute bolometric magnitude, $M_{\rm bol}$, plotted against
log\,$T_{\rm eff}$ for the four late M dwarfs in a) and a solid line
(some numbers on it are the values of stellar mass $M/M_{\odot}$)
is the theoretical HR diagram by \citet{Bar98}, reproduced
from figure 17 of Paper I. It is to be noted that the positions of
three objects GJ\,406, GJ\,644C, and GJ\,3849 are consistent with
 the theoretical HR diagram. 
For this reason, we apply these three objects to  revise the $M_{\rm 3.4} -
 {\rm log}\,T_{\rm eff}$  relation for late M dwarfs and we propose
a slightly modified $M_{\rm 3.4} - {\rm log}\,T_{\rm eff}$  relation
shown by a solid line in a).
}
\label{figure1}
\end{figure}

\begin{figure}
   \begin{center}
       \FigureFile(80mm,0mm){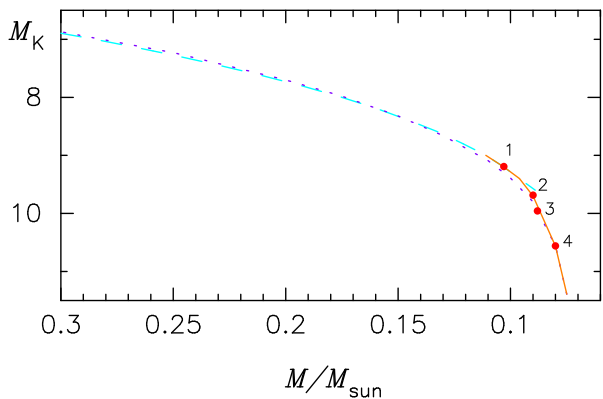}
   \end{center}
   \caption{Mass-luminosity ($M_{\rm K}$) relation by \citet{Del00} 
(dashed line) is extended to late M dwarfs using 
that based on the evolutionary models by \citet{Bar98}
(dotted line). The masses of the calibration stars in table 4 are 
obtained from the resulting mass-luminosity relation shown by a solid line.
}
\label{figure2}
\end{figure}

\begin{figure}
   \begin{center}
       \FigureFile(85.6mm,0mm){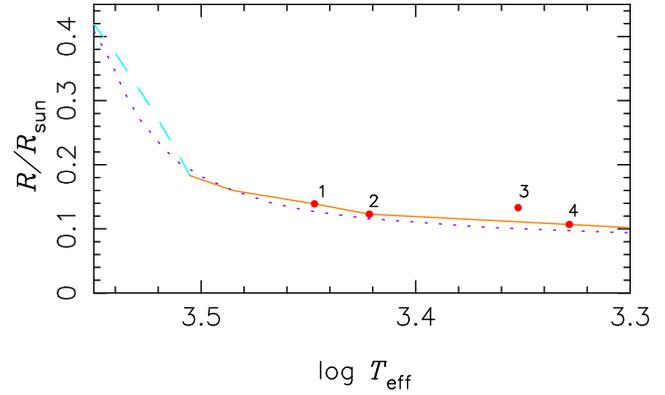}
   \end{center}
   \caption{$R/R_{\odot} - {\rm log}\,T_{\rm eff}$ relation by equation 8
of \citet{Boy12} (dashed line) is extended to late M dwarfs (solid line)  
based on our calibration stars no. 1, 2, \& 4 in table 4. The theoretical 
relation based on the evolutionary models by \citet{Bar98} is shown by 
a dotted line.
}\label{figure3}
\end{figure}

\begin{figure}
   \begin{center}
       \FigureFile(83.9mm,0mm){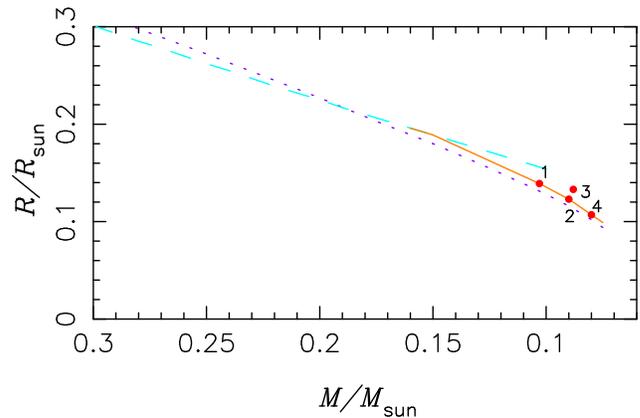}
   \end{center}
   \caption{$R/R_{\odot} - M/M_{\odot}$ relation by equation 10
of \citet{Boy12} (dashed line)  extended to late M dwarfs (solid line)
based on the calibration stars no. 1, 2, \& 4 in table 4.
A dotted line is based on the evolutionary model by \citet{Bar98}.
}
\label{figure4}
\end{figure}

\begin{figure}
   \begin{center}
       \FigureFile(85.25mm,0mm){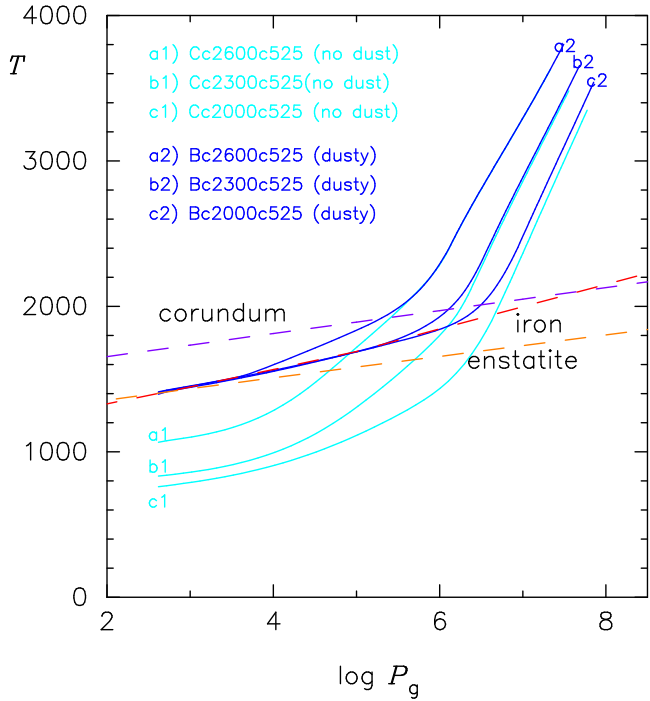}
   \end{center}
   \caption{Thermal structures of the dusty (curves a2, b2, and c2 shown by 
blue or black lines) and dust-free (curves a1, b1, and c1 shown by 
light sky or grey lines) models are compared for $T_{\rm eff}$ = 2600, 2300, 
and 2000\,K ({\it case c}, log\,$g$ = 5.25). The condensation temperatures 
of corundum, iron, and enstatite are  shown by dashed lines. 
}
\label{figure5}
\end{figure}

\begin{figure}
   \begin{center}
       \FigureFile(82.55mm,0mm){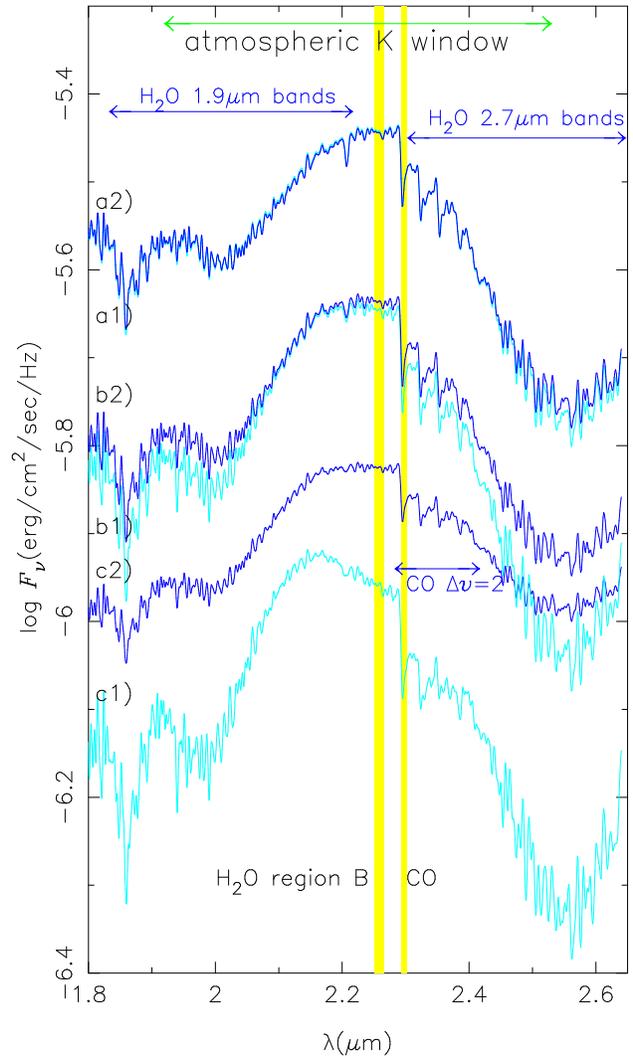}
   \end{center}
   \caption{Theoretical spectra ($R$ = 600) of the dusty (curves a2, b2, and 
c2 shown by blue or black lines) and dust-free (curves a1, b1, and c1 
shown by light sky or grey) models are compared for $T_{\rm eff}$ = 2600, 
2300, and 2000\,K ({\it case c}, log\,$g$ = 5.25). The effect of dust on 
the spectra is still negligible for $T_{\rm eff}$ = 2600, noticeable for 
$T_{\rm eff}$ = 2300, and quite  large for $T_{\rm eff}$ = 2000. The  regions 
indicated by CO and H$_2$O region B (shown by yellow or grey) are selected for
detailed analysis of CO and H$_2$O lines in 2MASS 1835+32. 
}
\label{figure6}
\end{figure}

\begin{figure}
   \begin{center}
       \FigureFile(70mm,0mm){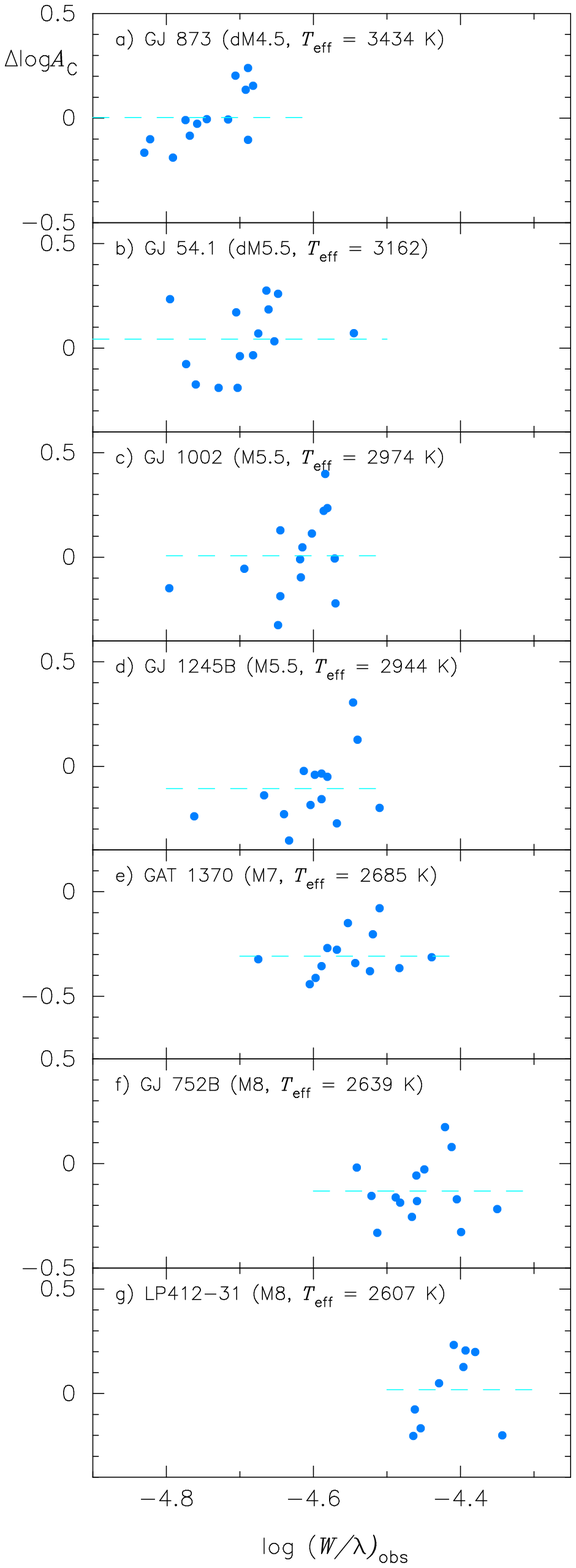}
   \end{center}
   \caption{The resulting logarithmic abundance corrections 
$\Delta$\,log\,$A_{\rm C}$ by the mini
curves-of-growth for the CO blends  plotted against the observed values of
log\,$(W/\lambda)_{\rm obs}$. A dashed line shows the mean correction.
a) GJ\,873, b) GJ\,54.1, c) GJ\,1002, d) GJ\,1245B, e) GAT\,1370, 
f) GJ\,752B, and g) LP\,412-31.    }
\label{figure7}
\end{figure}

\begin{figure}
   \begin{center}
   \FigureFile(72.75mm,0mm){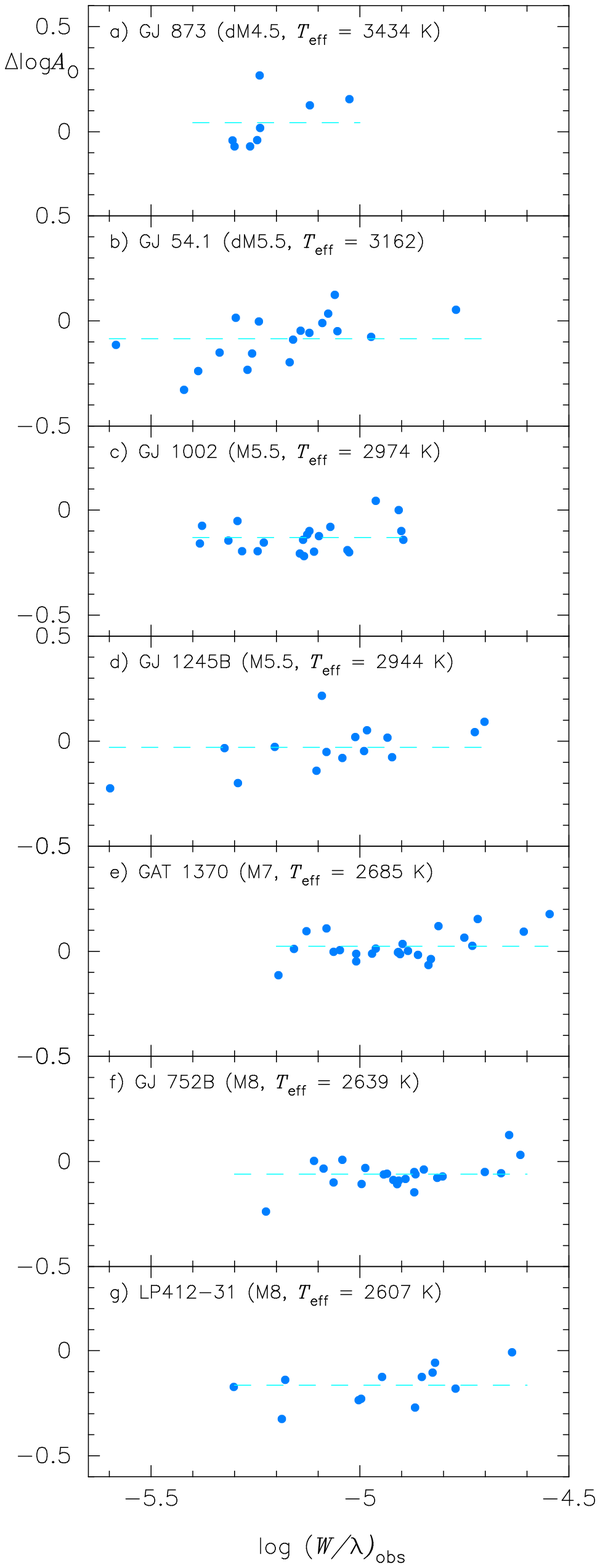}
   \end{center}
   \caption{The resulting logarithmic abundance corrections 
$\Delta$\,log\,$A_{\rm O}$ by the mini curves-of-growth 
for the H$_2$O blends in region B plotted against the observed values of
log\,$(W/\lambda)_{\rm obs}$. A dashed line shows the mean correction.
a) GJ\,873, b) GJ\,54.1, c) GJ\,1002, d) GJ\,1245B, e) GAT\,1370, 
f) GJ\,752B, and g) LP\,412-31.    }
\label{figure8}
\end{figure}

\begin{figure}
   \begin{center}
   \FigureFile(72.75mm,0mm){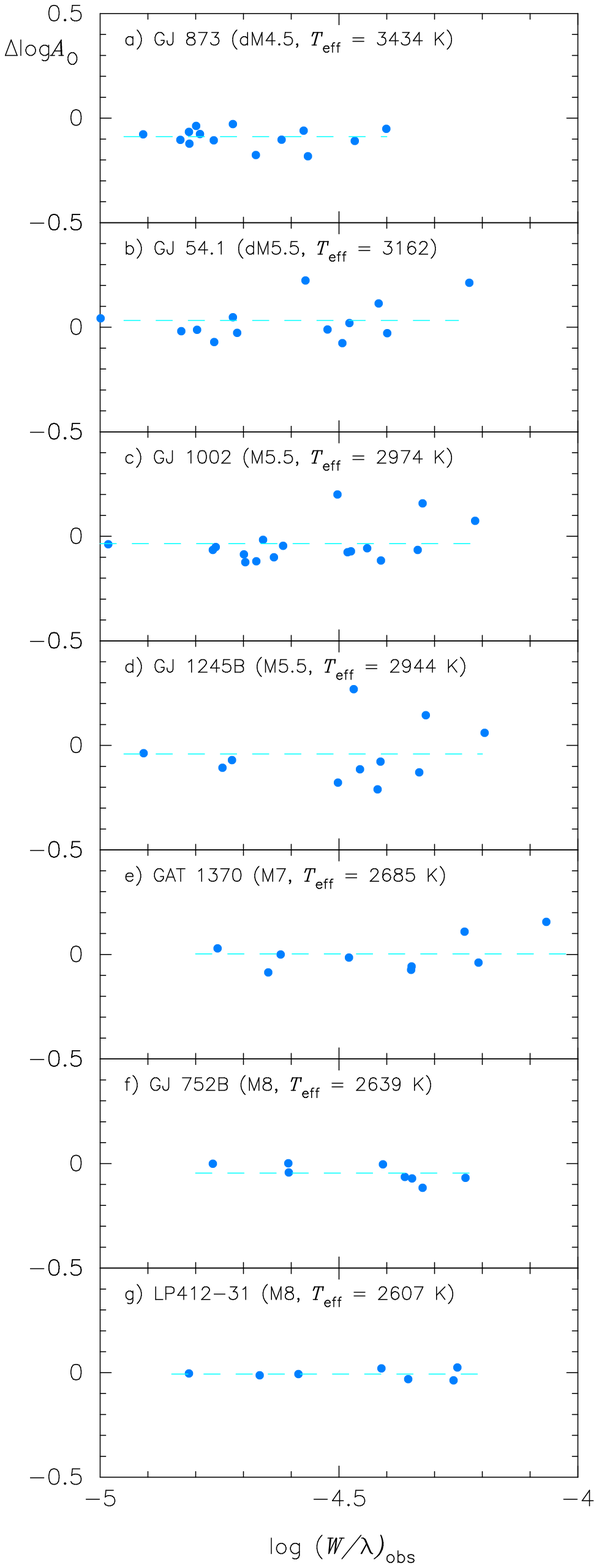}
   \end{center}
   \caption{The resulting logarithmic abundance corrections 
$\Delta$\,log\,$A_{\rm O}$ by the mini curves-of-growth 
for the H$_2$O blends in region A plotted against the observed values of
log\,$(W/\lambda)_{\rm obs}$. A dashed line shows the mean correction.
a) GJ\,873, b) GJ\,54.1, c) GJ\,1002, d) GJ\,1245B, e) GAT\,1370, 
f) GJ\,752B, and g) LP\,412-31.    }
\label{figure9}
\end{figure}

\onecolumn

\begin{figure}
   \begin{center}
       \FigureFile(145mm,0mm){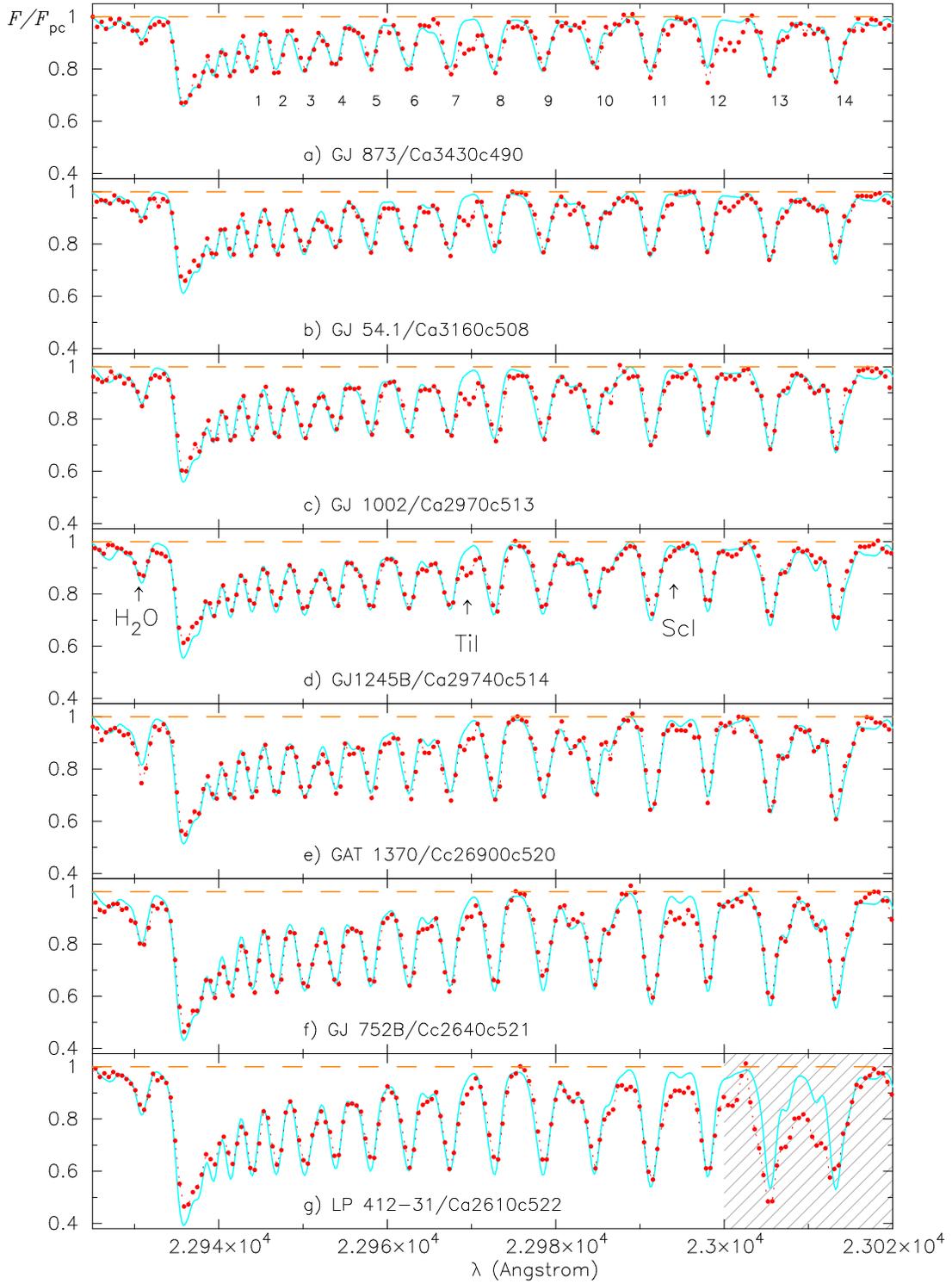}
   \end{center}
   \caption{
    Comparisons of the observed (filled circles) and predicted (solid line) 
CO spectra for the final carbon (table\,6) and oxygen (table\,8) 
abundances by the mini-CG method are shown for seven late M dwarfs:
a) GJ\,873, b) GJ\,54.1, c) GJ\,1002, d) GJ\,1245B, e) GAT\,1370, 
f) GJ\,752B, and g) LP\,412-31 (hatched region remains  unexplained). }
\label{figure10}
\end{figure}

\begin{figure}
   \begin{center}
       \FigureFile(145mm,0mm){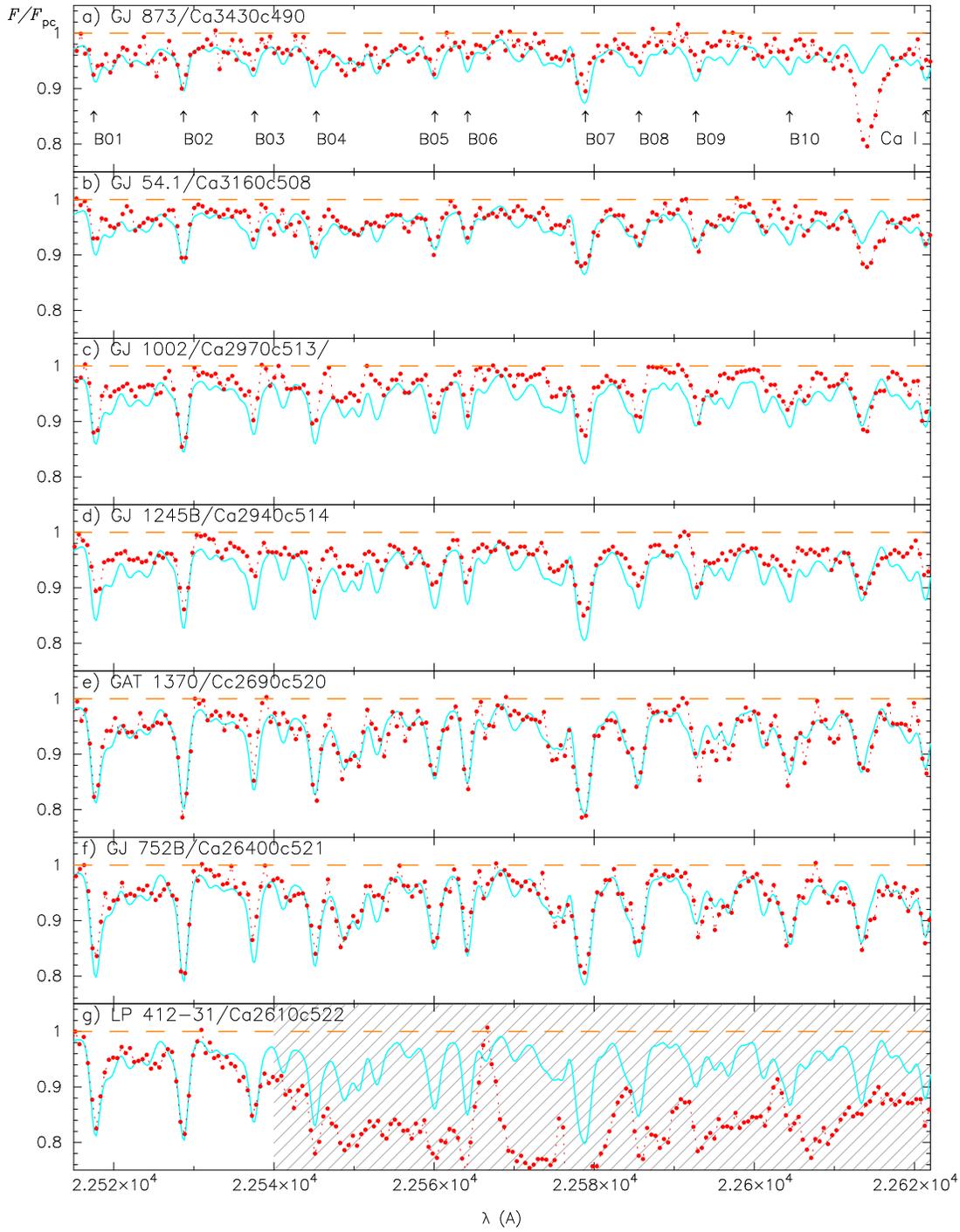}
   \end{center}
   \caption{
    Comparisons of the observed (filled circles) and predicted (solid line) 
H$_2$O spectra in region B for the final carbon (table\,6) and oxygen 
(table\,8) abundances by the mini-CG method are shown for seven 
late M dwarfs: a) GJ\,873, b) GJ\,54.1, c) GJ\,1002, d) GJ\,1245B, 
e) GAT\,1370, f) GJ\,752B, and g) LP\,412-31 (hatched region is
disturbed by unknown origin).
 }
\label{figure11}
\end{figure}

\begin{figure}
   \begin{center}
       \FigureFile(145mm,0mm){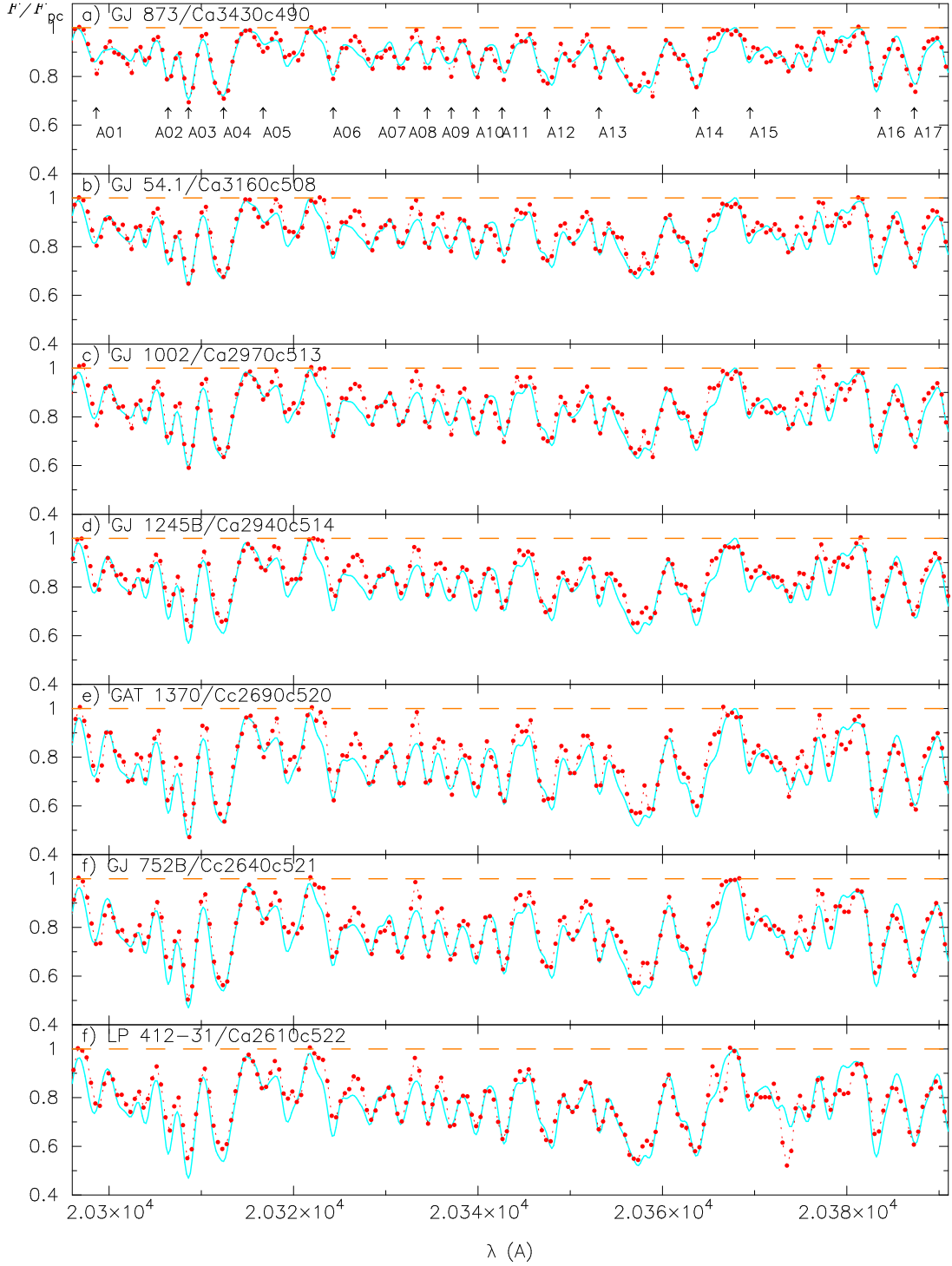}
   \end{center}
   \caption{
    Comparisons of the observed (filled circles) and predicted (solid line ) 
H$_2$O spectra in region A for the final carbon (table\,6) and oxygen 
(table\,10) abundances by the mini-CG method are shown for seven 
late M dwarfs: a) GJ\,873, b) GJ\,54.1, c) GJ\,1002, d) GJ\,1245B, 
e) GAT\,1370, f) GJ\,752B, and g) LP\,412-31.
 }
\label{figure12}
\end{figure}

\begin{figure}
   \begin{center}
       \FigureFile(145mm,0mm){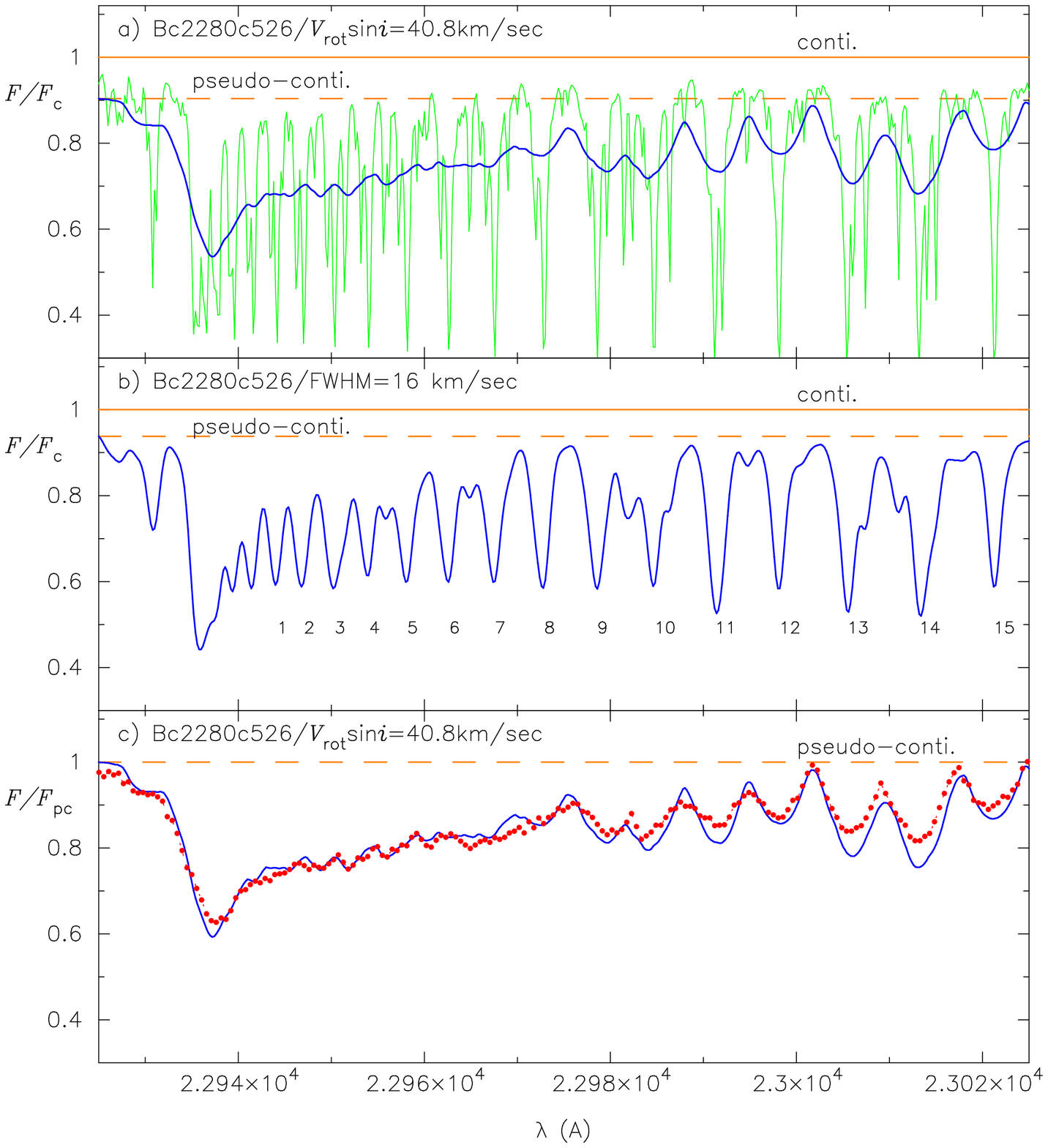}
   \end{center}
   \caption{a) The theoretical spectrum of CO (including the weak H$_2$O 
lines) based on the  dusty model Bc2280c526 evaluated with a sampling 
interval of 0.02\,\AA (thin line) is convolved with the rotation profile 
of $V_{\rm rot}{\rm sin}\,i = 40.8$\,km\,s$^{-1}$ (thick line).
b) For comparison, the same theoretical spectrum above is convolved with the
slit function (Gaussian) of FWHM = 16\,km\,s$^{-1}$. The numbers attached
are the ref. numbers for CO blends (table\,7 of Paper I and table 14
in Appendix 1). 
c) The observed spectrum of 2MASS 1835+32 (dots) is compared with the
theoretical spectrum broadened by  rotation (thick line
from a) above). Note that the theoretical spectrum is based on the
solar abundances of log\,$A_{\rm C} = -3.61 $ and log\,$A_{\rm O} = -3.31 $.
}
\label{figure13}
\end{figure}

\begin{figure}
   \begin{center}
       \FigureFile(145mm,0mm){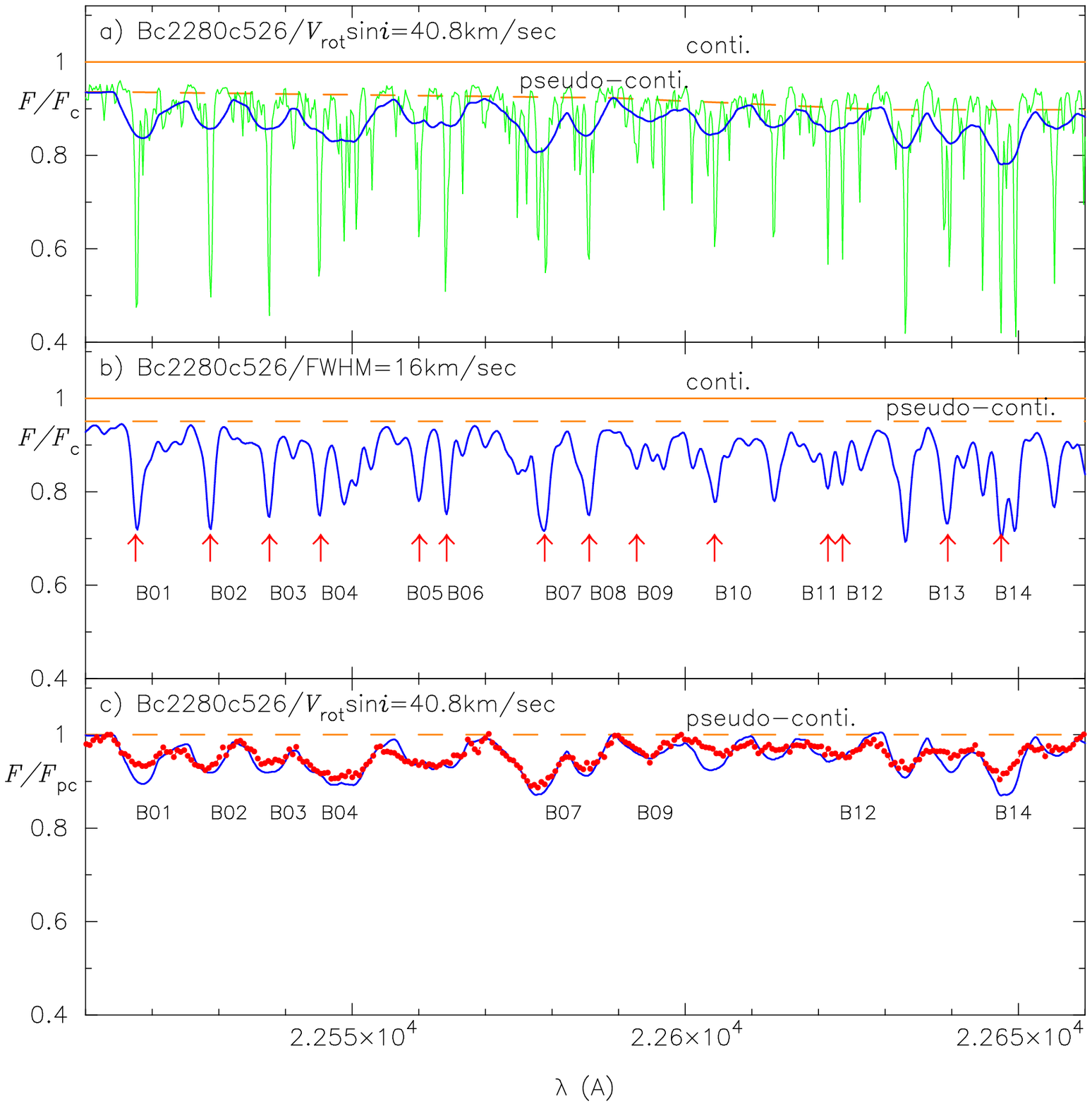}
   \end{center}
   \caption{a) The theoretical spectrum of  H$_2$O based on the  dusty
model Bc2280c526 evaluated with a sampling interval of 0.02\,\AA (thin
line) is convolved with the rotation profile of $V_{\rm rot}{\rm sin}\,i =
 40.8$\,km\,s$^{-1}$ (thick line).
b) For comparison, the same theoretical spectrum is convolved with the
slit function (Gaussian) of FWHM = 16\,km\,s$^{-1}$. The numbers attached
are the ref. numbers for H$_2$O blends (table\,4 of Paper II).
c) The observed spectrum of 2MASS 1835+32 (dots) is compared with the
theoretical spectrum broadened by rotation (thick line
from a) above). Note that the theoretical spectrum is based on the
solar abundances of log\,$A_{\rm C} = -3.61 $ and log\,$A_{\rm O} = -3.31 $.
}
\label{figure14}
\end{figure}

\twocolumn

\begin{figure}
   \begin{center}
       \FigureFile(72.75mm,0mm){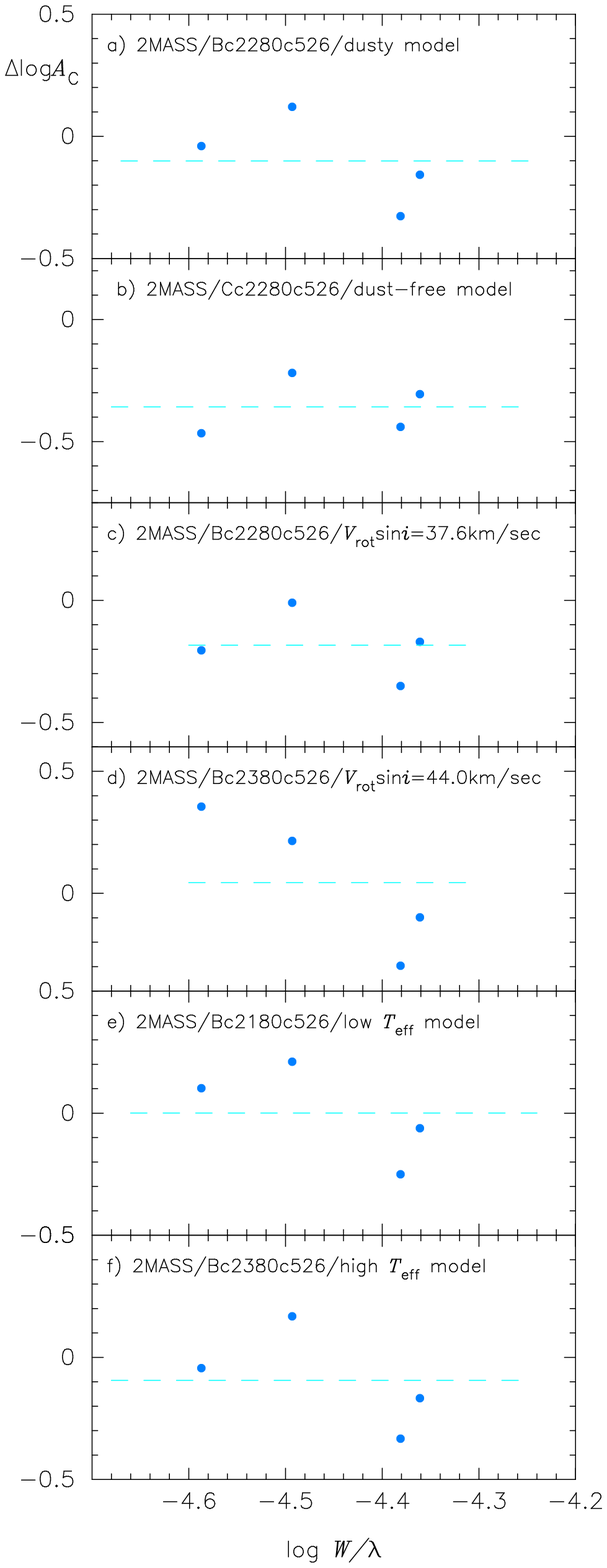}
   \end{center}
   \caption{The resulting logarithmic abundance corrections 
$\Delta$\,log\,$A_{\rm C}$  by the mini
curves-of-growth for the CO blends in 2MASS 1835+32
plotted against the observed values of log\,$(W/\lambda)_{\rm obs}$. 
A dashed line shows the mean correction.
a) Based on the dusty model  Bc2280c526 and $V_{\rm rot}{\rm sin}\,i =
 40.8$\,km\,s$^{-1}$.
b) Based on the dust-free model Cc2280c526 and $ V_{\rm rot}{\rm sin}\,i = 
40.8$\,km\,s$^{-1}$.
c) Based on the dusty model  Bc2280c526 and $ V_{\rm rot}{\rm sin}\,i = 
37.6$\,km\,s$^{-1}$ (i.e., lower $V_{\rm rot}{\rm sin}\,i$). 
d) Based  on the dusty model  Bc2280c526 and $ V_{\rm rot}{\rm sin}\,i = 
44.0$\,km\,s$^{-1}$ (i.e., higher $V_{\rm rot}{\rm sin}\,i$).
e) Based on the dusty model  Bc2180c528 (i.e., lower  $T_{\rm eff}$) 
and $V_{\rm rot}{\rm sin}\,i = 40.8$\,km\,s$^{-1}$.
f)  Based  on the dusty model  Bc2380c525 (i.e., higher  $T_{\rm eff}$)
and $V_{\rm rot}{\rm sin}\,i = 40.8$\,km\,s$^{-1}$.
}
\label{figure15}
\end{figure}

\begin{figure}
   \begin{center}
       \FigureFile(70mm,0mm){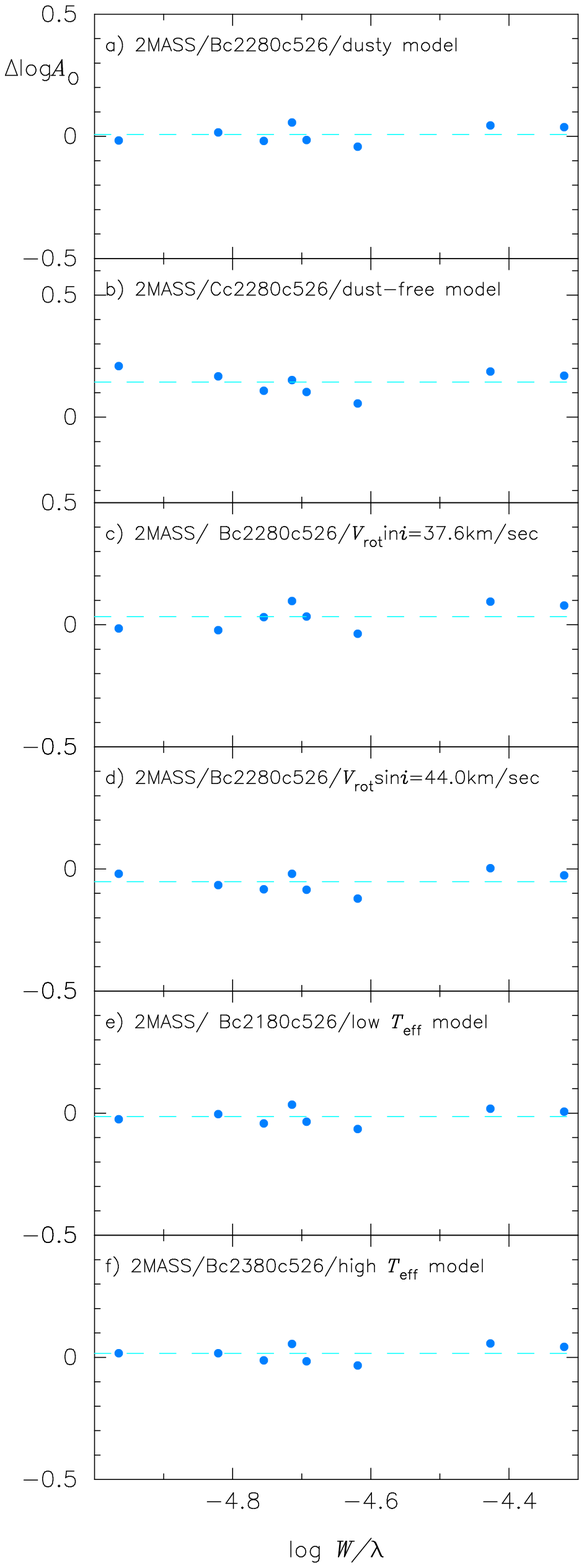}
   \end{center}
   \caption{The resulting logarithmic abundance corrections 
$\Delta$\,log\,$A_{\rm O}$ by the mini
curves-of-growth of the H$_2$O blends in 2MASS 1835+32 plotted 
against the observed values of log\,$(W/\lambda)_{\rm obs}$. A dashed 
line shows the mean correction.
a) Based on the dusty model  Bc2280c526 and $V_{\rm rot}{\rm sin}\,i =
 40.8$\,km\,s$^{-1}$.
b) Based on the dust-free model Cc2280c526 and $V_{\rm rot}{\rm sin}\,i = 
 40.8$\,km\,s$^{-1}$.
c) Based on the dusty model  Bc2280c526 and $ V_{\rm rot}{\rm sin}\,i = 
37.6$\,km\,s$^{-1}$. 
d) Based  on the dusty model  Bc2280c526 and $V_{\rm rot}{\rm sin}\,i = 
44.0$\,km\,s$^{-1}$. 
e) Based on the dusty model  Bc2180c528 
and $V_{\rm rot}{\rm sin}\,i = 40.8$\,km\,s$^{-1}$.
f)  Based  on the dusty model  Bc2380c525 
and  $V_{\rm rot}{\rm sin}\,i = 40.8$\,km\,s$^{-1}$.
}
\label{figure16}
\end{figure}

\onecolumn

\begin{figure}
   \begin{center}
       \FigureFile(145mm,0mm){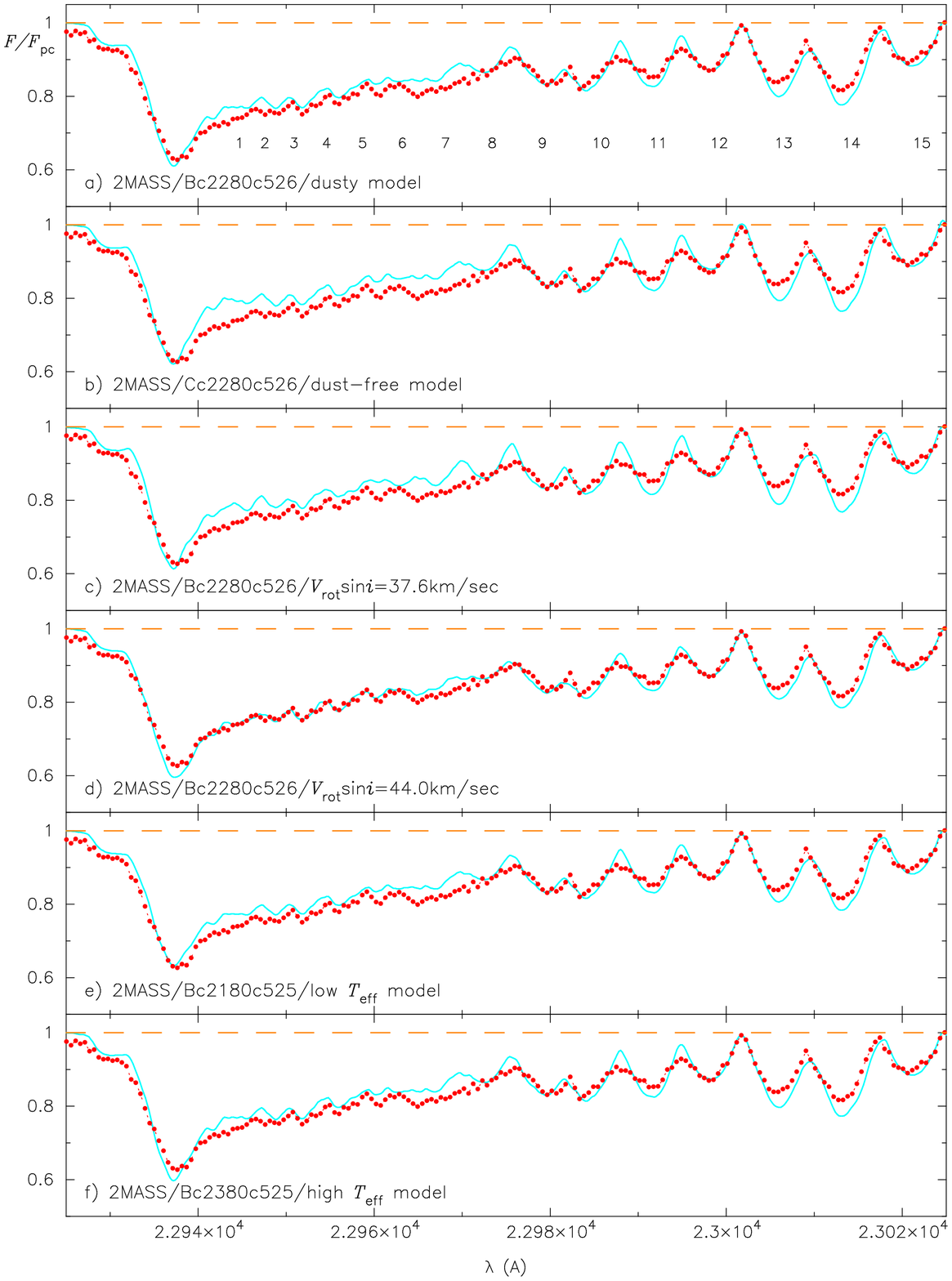}
   \end{center}
   \caption{ Comparison of the observed (dots) and predicted (solid line)
spectra of CO in 2MASS 1835+32 based on:
a) The dusty model Bc2280c526 with log\,$A_{\rm C}$ and log\,$A_{\rm O}$ in
table 12 (line no.\,1) and $V_{\rm rot}{\rm sin}\,i = 40.8$\,km\,s$^{-1}$. 
b) The dust-free model Cc2280c526 with log\,$A_{\rm C}$ and log\,$A_{\rm O}$ in
table 12 (line no.\,2) and $V_{\rm rot}{\rm sin}\,i = 40.8$\,km\,s$^{-1}$. 
 c) The dusty model Bc2280c526 with log\,$A_{\rm C}$ and log\,$A_{\rm O}$ in 
table 12 (line no.\,3) and $V_{\rm rot}{\rm sin}\,i = 37.6$\,km\,s$^{-1}$. 
 d) The dusty model Bc2280c526  with log\,$A_{\rm C}$ and log\,$A_{\rm O}$ in 
 table 12 (line no.\,4) and $V_{\rm rot}{\rm sin}\,i = 44.0$\,km\,s$^{-1}$. 
e) The dusty model Bc2180c528 with log\,$A_{\rm C}$ and log\,$A_{\rm O}$ in
table 12 (line no.\,5) and $V_{\rm rot}{\rm sin}\,i = 40.8$\,km\,s$^{-1}$.
f) The dusty model Bc2380c525 with log\,$A_{\rm C}$ and log\,$A_{\rm O}$ in 
table 12 (line no.\,6) and $V_{\rm rot}{\rm sin}\,i = 40.8$\,km\,s$^{-1}$.
}
\label{figure17}
\end{figure}

\begin{figure}
   \begin{center}
       \FigureFile(145mm,0mm){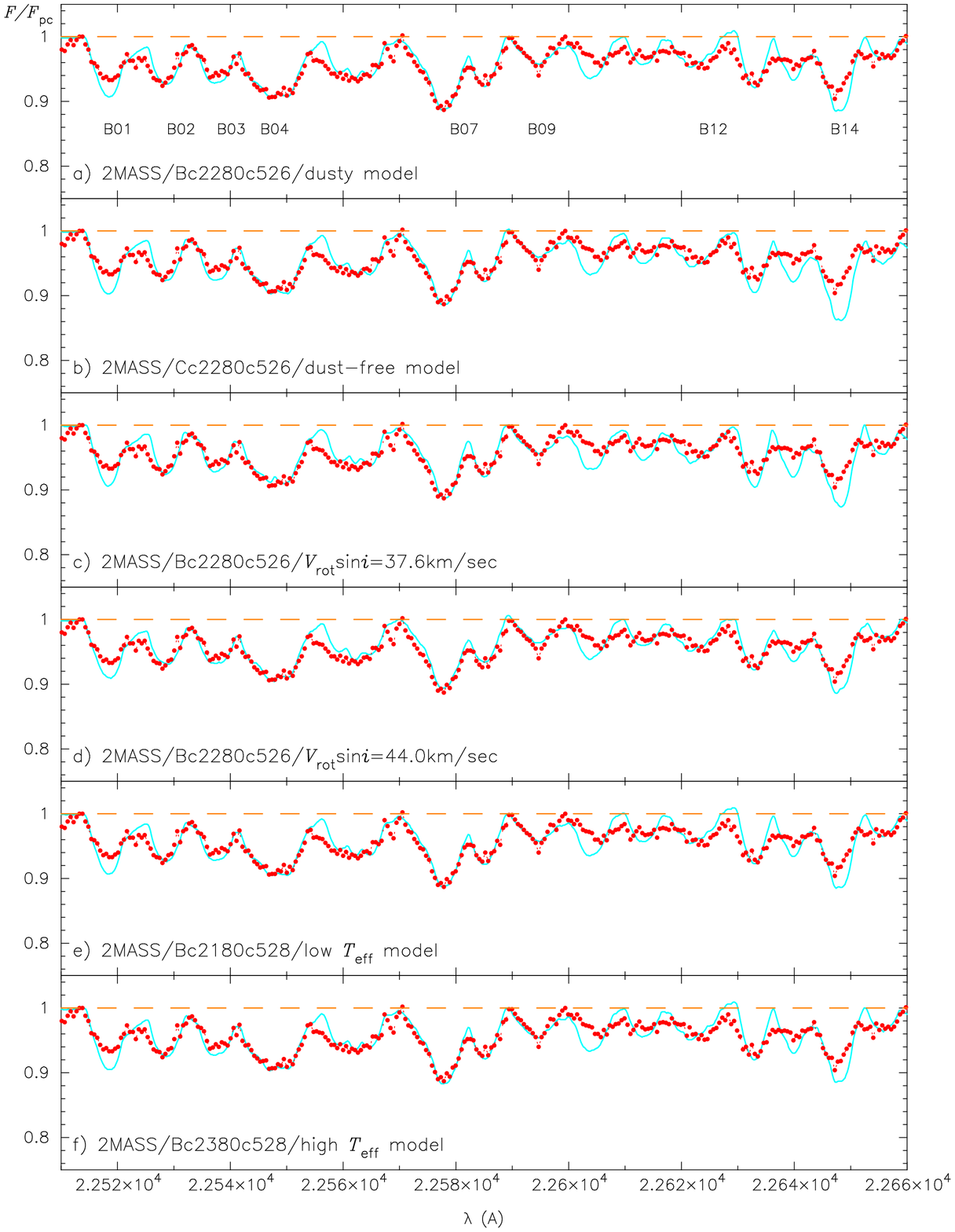}
   \end{center}
   \caption{Comparison of the observed (dots) and predicted (solid line)
spectra of H$_2$O in 2MASS 1835+32 based on:
a) The dusty model Bc2280c526 with log\,$A_{\rm C}$ and log\,$A_{\rm O}$ in
table 12 (line no.\,1) and $V_{\rm rot}{\rm sin}\,i = 40.8$\,km\,s$^{-1}$. 
b) The dust-free model Cc2280c526 with log\,$A_{\rm C}$ and log\,$A_{\rm O}$ in
table 12 (line no.\,2) and $V_{\rm rot}{\rm sin}\,i = 40.8$\,km\,s$^{-1}$. 
 c) The dusty model Bc2280c526 with log\,$A_{\rm C}$ and log\,$A_{\rm O}$ in 
table 12 (line no.\,3) and $V_{\rm rot}{\rm sin}\,i = 37.6$\,km\,s$^{-1}$. 
 d) The dusty model Bc2280c526  with log\,$A_{\rm C}$ and log\,$A_{\rm O}$ in 
 table 12 (line no.\,4) and $V_{\rm rot}{\rm sin}\,i = 44.0$\,km\,s$^{-1}$. 
e) The dusty model Bc2180c528 with log\,$A_{\rm C}$ and log\,$A_{\rm O}$ in
table 12 (line no.\,5) and $V_{\rm rot}{\rm sin}\,i = 40.8$\,km\,s$^{-1}$.
f) The dusty model Bc2380c525 with log\,$A_{\rm C}$ and log\,$A_{\rm O}$ in 
table 12 (line no.\,6) and $V_{\rm rot}{\rm sin}\,i = 40.8$\,km\,s$^{-1}$.
}
\label{figure18}
\end{figure}

\begin{figure}
   \begin{center}
       \FigureFile(120mm,0mm){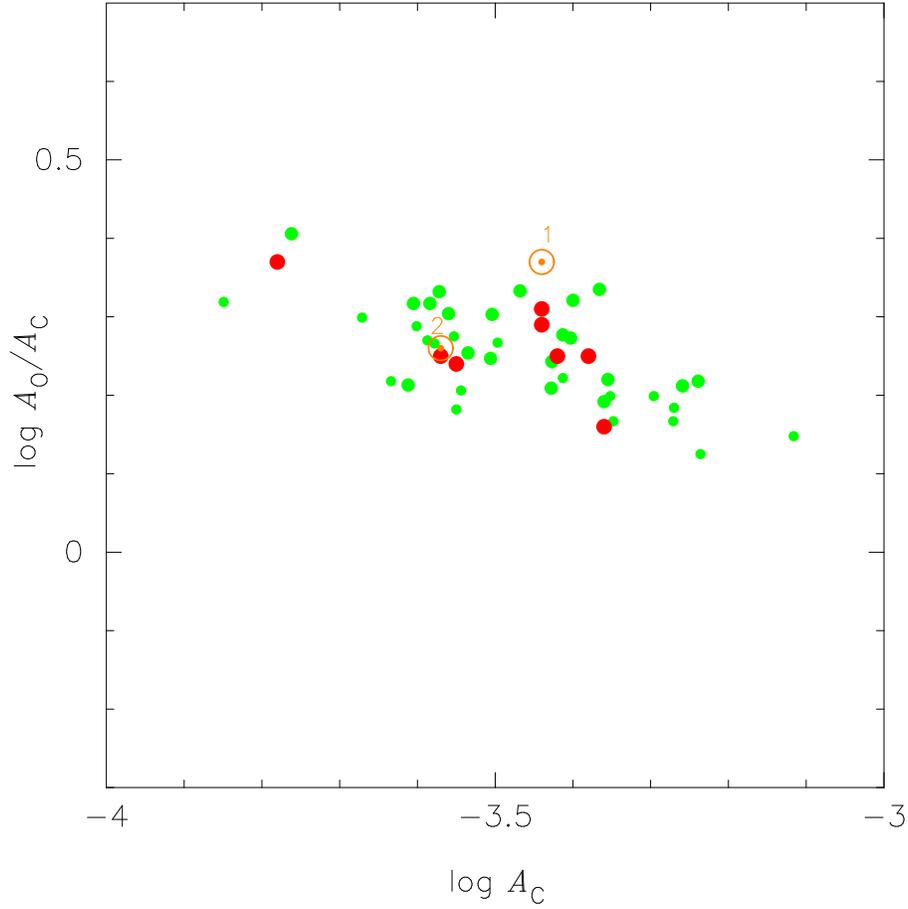}
   \end{center}
   \caption{The oxygen-to-carbon ratios in eight late M dwarfs
from table\,19 are plotted against the carbon abundances log\,$A_{\rm C}$
by red (or black) filled circles. Those in the early and middle M dwarfs 
from Paper II are shown by  green (or grey) filled circles. The solar 
values based on the classical \citep{And89} and more  recent 
result \citep{Asp09}  are also shown by $\odot$ mark with 1 and 2,
respectively.
}
\label{figure19}
\end{figure}

\begin{figure}
   \begin{center}
       \FigureFile(145mm,0mm){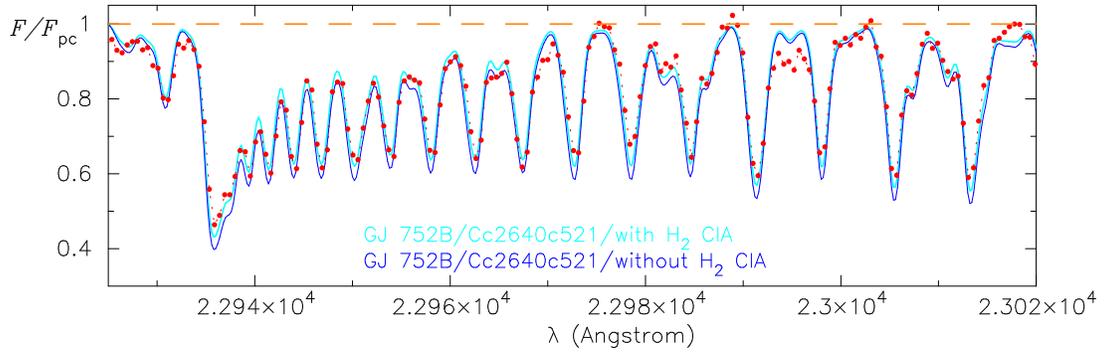}
   \end{center}
   \caption{The predicted spectrum of CO based on the photospheric model 
Cc2640c521 and log\,$A_{\rm C} = -3.550$ (table 6) but disregarding the
H$_2$ CIA (blue/black solid line) is compared with the one based on the
same model and log\,$A_{\rm C}$ but including the H$_2$ CIA 
(light sky/grey solid line). The observed spectrum of GJ\,752B is shown
by  filled circles.
}
\label{figure20}
\end{figure}

\begin{figure}
   \begin{center}
       \FigureFile(70mm,0mm){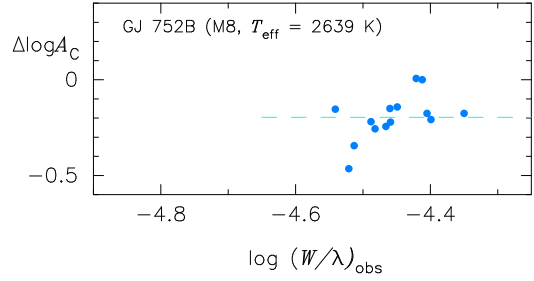}
   \end{center}
   \caption{
The logarithmic abundance corrections $\Delta {\rm log}\,A_{\rm C}$
to the value of log\,$A_{\rm C} = -3.550$ (based on the normal analysis 
of GJ\,752B including the H$_2$ CIA) needed to explain the observed 
EWs of CO blends by the use of the model 
photosphere Cc2640c521 disregarding the H$_2$ CIA.
The resulting $\Delta {\rm log}\,A_{\rm C} = -0.196 \pm 0.081$ implies that
the neglect of H$_2$ CIA results in an error of -0.20\,dex in 
log\,$A_{\rm C}$.
}
\label{figure21}
\end{figure}

\begin{figure}
   \begin{center}
       \FigureFile(120mm,0mm){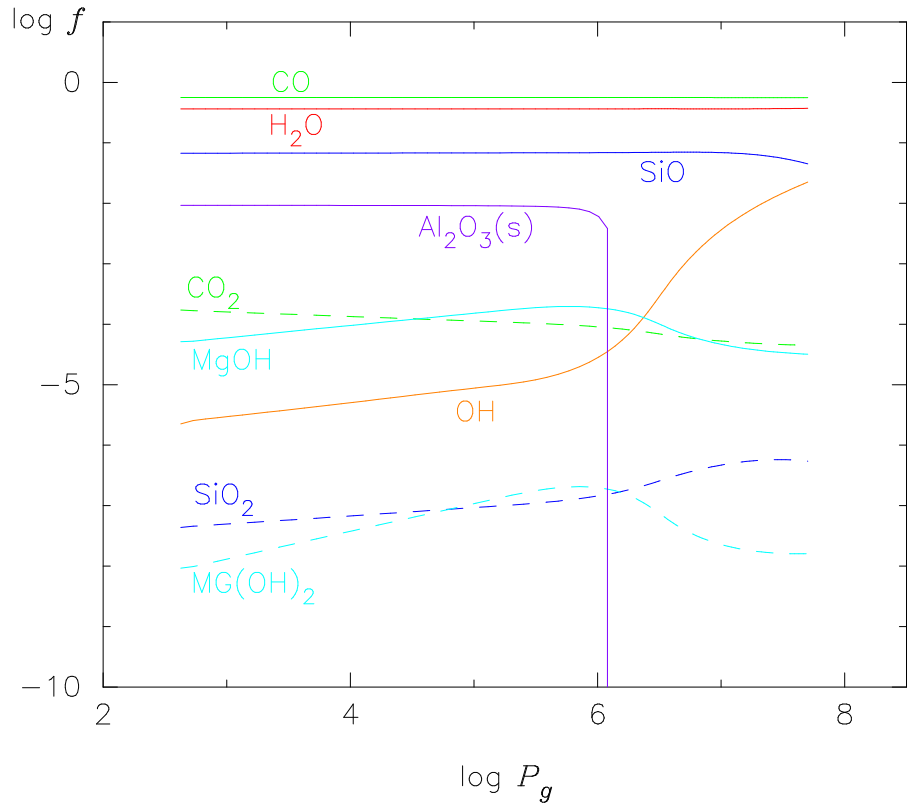}
   \end{center}
   \caption{
 Logarithm of the fraction $f$ (defined by equation 7 in the 
text) of oxygen atoms in the major molecules and grains
consuming oxygen  plotted against log\,$P_{\rm g}$ in the model 
photosphere Bc2280c526 of the M8.5 dwarf 2\,MASS 1835+32. Most oxygen
atoms are depleted into CO, H$_2$O, and SiO molecules, and about 1\%
of oxygen atoms are in corundum (Al$_2$O$_3$). Enstatite (MgSiO$_3$)
does not form in this model yet (see figure 5).
}
\label{figure22}
\end{figure}

\clearpage

\begin{table}
\begin{center}
\caption{Target stars.}
\label{table-targets}
\begin{tabular}{llcccc}
\hline
  Name                 & Other Name        & $K$ mag& SNR(28th) & SNR(25th)  & LGS\footnotemark[$*$] \\
\hline
GJ 54.1                &    YZ Cet        &    6.42 &  63   &  53    &    \\
GJ 752B                &      vB10        &    8.77 & 105   & 156    &    \\
GJ 873                 &      EV Lac      &  5.30   &  94   &  40    &    \\
GJ 1002                &    LHS 2         &    7.44 & 139   & 142    &    \\
GJ 1245B               &      LHS 3495    &    7.39 &  63   &  50    &    \\
GAT 1370               & Teegarden's star &   7.59  &  84   &  88    &   \\
LP 412-31              &                  &  10.64  &  31   &  66    & Y  \\
2MASS\,1835+32\footnotemark[$\dagger$] &    &   9.17  &  87   &  67    &  Y \\
\hline
\multicolumn{6}{l}{\hbox to 0pt{\parbox{80mm}{\footnotesize
   \par\noindent
   {}
   \par\noindent
   \footnotemark[$*$] A laser guide star was used, if Y.
   \par\noindent
   \footnotemark[$\dagger$] Abbreviation of 2MASSI\,J18353790+3259545.
   }}}
\end{tabular}
\end{center}
\end{table}

\begin{table}
\begin{center}
\caption{Ecehlle setting.}
\label{table-echelle}
\begin{tabular}{ccc}
\hline
  Order  &  Wavelength range (\AA) & Dispersion ({\AA} pixel$^{-1}$) \\
\hline
29th  & 19397$-$19867 & 0.460  \\
28th  & 20089$-$20575 & 0.475  \\
27th  & 20833$-$21335 & 0.491  \\
26th  & 21634$-$22154 & 0.508  \\
25th  & 22500$-$23039 & 0.527  \\
24th  & 23437$-$23997 & 0.547  \\
23rd  & 24456$-$25038 & 0.569  \\
\hline
\end{tabular}
\end{center}
\end{table}

\begin{table*}
\begin{center}
\caption{Fundamental  parameters of the program stars.}
  \begin{tabular}{llccccccc}
  \hline
obj. &  sp.type\footnotemark[$*$] &  $p$(msec)\footnotemark[$\dagger$] &  $F_{\rm 3.4}$(mag)
\footnotemark[$\ddagger$] &  $M_{\rm 3.4}$(mag)\footnotemark[$\S$]  &  $T_{\rm eff}$\footnotemark[$\|$]  & 
$R/R_{\odot}$\footnotemark[$\#$] & $M/M_{\odot}$\footnotemark[$**$]  & log\,$g$   \\
\hline
GJ\,54.1 & dM5.5e & 271.010 $\pm$ 8.360 & 6.167 $\pm$ 0.044 & 8.33 $\pm$ 0.11 & 3162. &  0.177 &  0.139 & 5.084 \\    
GJ\,752B &  M8     & 170.360 $\pm$ 1.000 & 8.317 $\pm$ 0.020 & 9.47 $\pm$ 0.03 & 2639. & 0.123 & 0.090 & 5.213 \\
GJ\,873  &  dM4.5e & 195.220 $\pm$ 1.870 & 5.241 $\pm$ 0.063 & 6.69 $\pm$ 0.08 & 3434. & 0.353 & 0.363 & 4.903 \\
GJ\,1002 &  M5.5   & 213.000 $\pm$ 3.600 & 7.176 $\pm$ 0.028 & 8.82 $\pm$ 0.06 & 2974. & 0.153 & 0.116 & 5.133 \\
GJ\,1245B & M5.5   & 220.000 $\pm$ 1.000 & 7.178 $\pm$ 0.066 & 8.89 $\pm$ 0.08 & 2944. & 0.151 & 0.114 & 5.138 \\
GAT\,1370 & M8/9   & 260.630 $\pm$ 2.690 & 7.322 $\pm$ 0.027 & 9.40 $\pm$ 0.05 & 2685. & 0.128 & 0.094 & 5.199 \\
LP\,412-31 & M8    &  68.300 $\pm$ 0.600 & 10.352 $\pm$ 0.023 & 9.52 $\pm$ 0.04 & 2607. & 0.122 & 0.089 & 5.217 \\
2MASS\,1835+32 & M8.5 & 176.5 $\pm$ 0.5 & 8.837 $\pm$ 0.020 & 10.07 $\pm$ 0.03 & 2275. & 0.112 & 0.083 & 5.261 \\
  \hline
  \multicolumn{9}{@{}l@{}}{\hbox to 0pt{\parbox{170mm}{\footnotesize
      \hbox{}
      \par\noindent
      \footnotemark[$*$]  dM by \citet{Joy74}, and M by SIMBAD.
      \par\noindent
      \footnotemark[$\dagger$] By Hipparcos \citep{Lee07}, except for 2MASS\,1835+32 by RECONS.
      \par\noindent
      \footnotemark[$\ddagger$] $W1$ band flux centered at 3.4\,$\micron$ from the WISE All-Sky Release for $F_{\rm 3.4} 
        < 8.0$\,mag and from the AllWISE catalog for $F_{\rm 3.4} > 8.0$\,mag \citep{Wri10}.
      \par\noindent
      \footnotemark[$\S$] Absolute magnitude at 3.4\,$\micron$ based on $F_{\rm 3.4}$.
      \par\noindent
      \footnotemark[$\|$] Based on $M_{\rm 3.4} - {\rm log}\,T_{\rm eff}$ 
relation shown by the dashed line in figure\,1 of Paper I for $M_{\rm 3.4} < 9.0 $ and based on the solid line in 
figure\,1a for $M_{\rm 3.4} > 9.0 $.
      \par\noindent
      \footnotemark[$\#$] Based on radius $-$ log\,$T_{\rm eff}$ relation of equation 8 by \citet{Boy12} for dM4.5 $-$ dM6, and 
        on the solid line in figure\,3 for dM7 $-$ dM8.5.
      \par\noindent
      \footnotemark[$**$] Based on mass $-$ radius relation of equation 10 by \citet{Boy12} and on the solid line in figure\,4.
      }\hss}}
  \end{tabular}
\end{center}
\end{table*}

\begin{table*}
\begin{center}
\caption{Fundamental  parameters of the calibration objects.}
\small 
  \begin{tabular}{lllcccccccccc}
  \hline
no. & obj. & other\,name & $p$(msec)\footnotemark[$*$] &  
$F_{\rm 3.4}$(mag)\footnotemark[$\dagger$] &  
$M_{\rm 3.4}$(mag)\footnotemark[$\ddagger$] & $T_{\rm eff}$\footnotemark[$\S$]  &  log\,$f_{\rm bol}$\footnotemark[$\|$] &
$log\,L/L_{\odot}$\footnotemark[$\#$] & $R/R_{\odot}$\footnotemark[$**$]  
& $K$\footnotemark[$\dagger\dagger$] &
$M_{\rm K}$\footnotemark[$\ddagger\ddagger$] &  
$M/M_{\odot}$\footnotemark[$\S\S$]  \\
\hline
1 & GJ\,406 & Wolf\,359   &  419.1 &  5.807 &  8.92 & 2800. &  -8.221 &  -2.975 & 0.139 &  6.08 &  9.19 & 0.103 \\
2 & GJ\,644C & vB\,8  &  148.92 & 8.619 &  9.48 & 2640. &  -9.329 & -3.184 &  0.123 &  8.82 &  9.69 & 0.090 \\
3 & GJ\,752B & vB\,10  &  170.36 & 8.317 &  9.47 & 2250. &  -9.422 & -3.394 & 0.133 &  8.80 &  9.96 & 0.088 \\
4 & GJ\,3849 & LHS\,2924 &  95.0  & 10.428 & 10.32 & 2130. & -10.210 &   -3.675 & 0.107 & 10.67 & 10.56 & 0.080 \\
\hline 
  \multicolumn{12}{@{}l@{}}{\hbox to 0pt{\parbox{180mm}{\footnotesize
      \hbox{}
      \par\noindent
      \footnotemark[$*$] Parallax by RECONS for GJ\,406 and GJ\,644C, by Hipparcos \citep{Lee07} for GJ\,752B, and
            by \citet{Gli91} for GJ\,3849.
      \par\noindent
        \footnotemark[$\dagger$]$W1$ band flux centered at 3.4\,$\micron$ from the WISE All-Sky Release for $F_{\rm 3.4} 
        < 8.0$\,mag and from the AllWISE catalog for $F_{\rm 3.4} > 8.0$\,mag \citep{Wri10}.
      \par\noindent
      \footnotemark[$\ddagger$] Absolute magnitude at 3.4\,$\micron$ based on $F_{\rm 3.4}$.  
      \par\noindent
      \footnotemark[$\S$] Based on the infrared flux method \citep{Tsu96a}.
      \par\noindent
      \footnotemark[$\|$] Observed bolometric flux based on the integration of SED \citep{Tsu96a}.
      \par\noindent
      \footnotemark[$\#$] Bolometric luminosity based on the third and seventh columns. 
      \par\noindent
      \footnotemark[$**$] Based on $L/L_{\odot}$ and $T_{\rm eff}$.
      \par\noindent
      \footnotemark[$\dagger\dagger$] Observed $K$ magnitude \citep{Leg92}.
      \par\noindent
      \footnotemark[$\ddagger\ddagger$] Absolute magnitude at $K$.
      \par\noindent
      \footnotemark[$\S\S$] Based on $M_{\rm K} - M/M_{\odot}$ of \citet{Del00} extended by the use of  
          $M_{\rm K} - M/M_{\odot}$ from the evolutionary models of \citet{Bar98}. see figure 2.
      }\hss}}
  \end{tabular}
\end{center}
\end{table*}

\begin{table*}
\begin{center}
\caption{log\,$(W/\lambda)_{\rm obs}$\footnotemark[$*$] of 
CO blends (2-0 band) in 8 late M dwarfs.}
  \begin{tabular}{ccccccccc}
  \hline
 Ref. no\footnotemark[$\dagger$] & GJ\,54.1 & GJ\,752B & GJ\,873 & GJ\,1002 
& GJ\,1245B & GAT\,1370 & LP\,412-31 & 2MASS\,1835+32 \\ 
\hline
      1  & -4.705 & -4.460 & -4.774 & -4.645 & -4.613 & -4.581 & -4.393 &   ---  \\
      2  & -4.661 & -4.459 & -4.758 & -4.615 & -4.581 & -4.519 & -4.454 &   --- \\
      3  & -4.545 & -4.350 & -4.689 & -4.570 & -4.510 & -4.439 & -4.343 &   --- \\
      4  & -4.664 & -4.421 & -4.706 & -4.584 & -4.546 & -4.553 & -4.409 &   --- \\
      5  & -4.648 & -4.449 & -4.830 & -4.602 & -4.540 & -4.589 & -4.429 &   --- \\
      6  & -4.700 & -4.482 & -4.768 & -4.618 & -4.640 & -4.597 & -4.396 &   --- \\
      7  & -4.675 & -4.412 & -4.692 & -4.581 & -4.598 & -4.510 & -4.380 &   --- \\
      8  & -4.760 & -4.488 & -4.689 & -4.586 & -4.589 & -4.568 & -4.462 &   --- \\
      9  & -4.729 & -4.513 & -4.791 & -4.645 & -4.604 & -4.605 & -4.464 &   --- \\
     10  & -4.773 & -4.521 & -4.822 & -4.694 & -4.667 &  ---   &   ---  &   --- \\
     11  & -4.653 & -4.405 & -4.682 & -4.571 & -4.633 & -4.523 &   ---  &   --- \\
     12  & -4.795 & -4.541 &  ---   & -4.796 & -4.762 & -4.675 &   ---  & -4.493 \\
     13  & -4.682 & -4.466 & -4.745 & -4.617 & -4.589 & -4.543 &   ---  & -4.361 \\
     14  & -4.703 & -4.399 & -4.716 & -4.648 & -4.568 & -4.483 &   ---  & -4.381 \\
     15  &  ---   &  ---   &  ---   &  ---   &  ---   &  ---   &   ---  & -4.587 \\
\hline
  \multicolumn{9}{@{}l@{}}{\hbox to 0pt{\parbox{150mm}{\footnotesize
      \hbox{}
      \par\noindent
      \footnotemark[$*$] The equivalent width $W$ is measured by referring to the 
        pseudo-continuum.
      \par\noindent
      \footnotemark[$\dagger$]  Ref. nos. defined in table 7 and figure\,4 of Paper I,
        except for ref. no.15 defined in table 14 (Appendix 1) and figure 13b.
      }\hss}}
  \end{tabular}
\end{center}
\end{table*}

\begin{table*}
\begin{center}
\caption{Carbon abundances from the CO lines.}
  \begin{tabular}{lcclcccccrr}
  \hline
obj. & $T_{\rm eff}$ & log\,$g$ & model1\footnotemark[$*$] & $\Delta$\,log\,$A_{\rm C}^{(1)}$\footnotemark[$\dagger$]
 & log\,$A_{\rm C}^{(1)}$ & model2\footnotemark[$\ddagger$] & $\Delta$\,log\,$A_{\rm C}^{(2)}$\footnotemark[$\S$] & 
log\,$A_{\rm C}^{(2)}$ & $N$\footnotemark[$\|$] &  $\chi^{2}$\footnotemark[$\#$] \\
\hline
GJ\,54.1 &  3162. & 5.084 &  Ca3200c50 & +0.042 & -3.358 $\pm$ 0.112 & Ca3160c508 & -0.022 &  -3.380 $\pm$ 0.086 & 14 & 1.511  \\
GJ\,752B &  2639. & 5.213 &  Ca2600c525 & -0.131 & -3.531 $\pm$ 0.098 & Cc2640c521 & -0.019 & -3.550 $\pm$ 0.066 & 14 & 20.114 \\ 
GJ\,873  &  3434. & 4.903 & Ca3400c50  & -0.003 & -3.397 $\pm$ 0.094 & Ca3430c490 & -0.043 & -3.440 $\pm$ 0.092 & 13 & 1.474   \\
GJ\,1002  & 2974. & 5.133 & Ca3000c50  & +0.007 & -3.393 $\pm$ 0.135 & Ca2970c513 & -0.024 & -3.417 $\pm$ 0.088 & 14 & 14.710  \\
GJ\,1245B & 2944. & 5.138 & Ca2900c525 & -0.107 & -3.507 $\pm$ 0.116 & Ca2940c514 & +0.072 & -3.435 $\pm$ 0.098 & 14 &  1.372  \\   
GAT\,1370 & 2685. & 5.199 & Ca2700c525 & -0.301 & -3.701 $\pm$ 0.071 & Cc2690c520 & -0.080 & -3.781 $\pm$ 0.061 & 13 &  5.716  \\
LP\,412-31 & 2607. & 5.217 & Ca2600c525 & +0.019 & -3.381 $\pm$ 0.123 & Ca2610c522 & +0.022 & -3.357 $\pm$ 0.068 & 9 & 17.296  \\ 
\hline
  \multicolumn{11}{@{}l@{}}{\hbox to 0pt{\parbox{150mm}{\footnotesize
      \hbox{}
      \par\noindent
      \footnotemark[$*$] Model photosphere from the UCM grid.
      \par\noindent
      \footnotemark[$\dagger$] First correction to the assumed values for the model of the Ca series; 
         log\,$A_{\rm C}^{(0)} = -3.40 $ and log\,$A_{\rm O}^{(0)} = -3.08 $. 
      \par\noindent
      \footnotemark[$\ddagger$] Specified model for $T_{\rm eff}$ and log\,$g$ in the second and third columns, respectively.
      \par\noindent
      \footnotemark[$\S$] Second correction to the starting values  log\,$A_{\rm C}$ = log\,$A_{\rm C}^{(1)}$ in the sixth column 
         of table 6 and log\,$A_{\rm O}$ = log\,$A_{\rm O}^{(1)}$ in the fourth column of table 8.
      \par\noindent
      \footnotemark[$\|$]  Number of CO blends used for the mini-CG analysis. 
      \par\noindent
      \footnotemark[$\#$]  $\chi^{2}$ value for the comparison of the observed and  predicted spectra in figure\,10.
      }\hss}}
  \end{tabular}
\end{center}
\end{table*}

\begin{table*}
\begin{center}
\caption{log\,$(W/\lambda)_{\rm obs}$\footnotemark[$*$] of 
H$_2$O blends in region B for 8 late M dwarfs.}
  \begin{tabular}{ccccccccc}
  \hline
 Ref. no\footnotemark[$\dagger$] & GJ\,54.1 & GJ\,752B & GJ\,873 & GJ\,1002 
& GJ\,1245B & GAT\,1370 & LP\,412-31 & 2MASS\,1835+32 \\ 
\hline
    B01 &  ---   & -4.870 &  ---   & -5.026 &  ---   & -4.830 & -4.868 & -4.693 \\
    B02 &  ---   & -5.225 &  ---   &  ---   &  ---   &  ---   & -5.187 & -4.714 \\
    B03 & -5.336 & -4.911 &  ---   & -5.121 & -5.011 & -4.885 & -4.852 & -4.755 \\
    B04 & -5.142 & -4.870 & -5.120 & -4.962 & -4.983 & -4.812 & -4.820 & -4.320  \\
    B05 & -5.387 &  ---   & -5.300 &  ---   &  ---   & -5.195 &  ---   & ---     \\
    B06 & -5.168 & -4.867 &  ---   & -5.134 & -5.042 & -4.908 &  ---   & ---     \\
    B07 & -5.121 & -4.943 &  ---   & -5.126 &  ---   & -4.962 &  ---   & -4.427  \\
    B08 & -5.242 & -5.063 &  ---   & -5.315 & -5.204 & -5.048 &  ---   & ---     \\
    B09 & -4.770 & -4.662 &  ---   & -4.896 & -4.725 & -4.608 &  ---   & -4.965  \\
    B10 &  ---   &  ---   &  ---   & -5.030 &  ---   &  ---   &  ---   & ---     \\
    B11 &  ---   &  ---   &  ---   &  ---   &  ---   &  ---   &  ---   & ---     \\
    B12 & -5.421 &  ---   &  ---   &  ---   &  ---   & -4.904 &  ---   & -4.821  \\
    B13 &  ---   &  ---   &  ---   &  ---   &  ---   &  ---   &  ---   & ---     \\
    B14 &  ---   & -4.987 &  ---   & -5.282 & -5.292 & -5.063 &  ---   & -4.619  \\
    B15 &  ---   & -5.087 &  ---   & -5.383 & -5.598 & -5.158 &  ---   & ---     \\
    B16 & -5.269 & -4.815 & -5.263 & -5.110 & -5.104 & -4.750 & -4.636 &  ---    \\
    B17 & -5.054 & -4.802 & -5.246 & -5.144 & -4.990 & -4.861 &  ---   &  ---    \\
    B18 &  ---   & -4.907 &  ---   & -5.245 &  ---   & -5.009 &  ---   &  ---    \\
    B19 & -5.584 & -5.042 & -5.240 & -5.378 & -5.091 & -5.080 & -5.179 & ---     \\
    B20 & -5.090 & -4.920 & -5.239 & -5.071 & -5.080 & -4.898 & -5.003 & ---　　\\　　
    B21 & -5.060 & -4.935 & ---    & -5.230 &  ---   & -5.009 & -4.947 & ---    \\
    B22 &  ---   &  ---   &  ---   &  ---   &  ---   &  ---   &  ---   & ---    \\
    B23 & -5.258 & -4.996 &  ---   & -5.136 &  ---   & -4.971 &  ---   & ---    \\
    B24 & -5.076 & -4.847 &  ---   &  ---   &  ---   &  ---   &  ---   & ---     \\
    B25 &  ---   &  ---   &  ---   &  ---   &  ---   &  ---   &  ---   & ---     \\
    B26 & -5.160 & -4.891 & -5.305 & -5.098 &  ---   & -4.909 & -4.997 & ---     \\
    B27 &  ---   &  ---   &  ---   &  ---   &  ---   & -4.836 &  ---   & ---    \\
\hline
  \multicolumn{9}{@{}l@{}}{\hbox to 0pt{\parbox{150mm}{\footnotesize
      \hbox{}
      \par\noindent
      \footnotemark[$*$] The equivalent width $W$ is measured by referring to the 
        pseudo-continuum.
      \par\noindent
      \footnotemark[$\dagger$]  Ref. no. defined in table 4 and figure\,3 of Paper II.
      }\hss}}
  \end{tabular}
\end{center}
\end{table*}

\begin{table*}
\begin{center}
\caption{Oxygen abundances log\,$A_{\rm O}^{B}$ from the H$_2$O blends in region B.}
  \begin{tabular}{lcccccrr}
  \hline
obj.  & model2\footnotemark[$*$] & $\Delta$\,log\,$A_{\rm O}^{(1)}$\footnotemark[$\dagger$]
 & log\,$A_{\rm O}^{(1)}$  & $\Delta$\,log\,$A_{\rm O}^{(2)}$\footnotemark[$\ddagger$] & 
log\,$A_{\rm O}^{(2)}$ & $N$\footnotemark[$\S$] &  $\chi^{2}$\footnotemark[$\|$] \\
\hline
GJ\,54.1 &   Ca3160c508 &    -0.085 & -3.193 $\pm$ 0.079 & +0.050 &  -3.143 $\pm$ 0.049 &  18  &  1.344  \\   
GJ\,752B &   Cc2640c521 &    -0.060 & -3.291 $\pm$ 0.046 & -0.010 &  -3.301 $\pm$ 0.042 &  23  & 13.946  \\
GJ\,873  &   Ca3430c490 &    +0.043 & -3.054 $\pm$ 0.085 & -0.015 &  -3.069 $\pm$ 0.103 &   8  &  2.154  \\
GJ\,1002 &   Ca2970c513 &    -0.131 & -3.224 $\pm$ 0.047 & +0.071 &  -3.153 $\pm$ 0.036 &  21  & 18.918  \\
GJ\,1245B &  Ca2940c514 &    -0.029 & -3.236 $\pm$ 0.076 & +0.104 &  -3.132 $\pm$ 0.050 &  15  &  2.479  \\
GAT\,1370 &  Cc2690c520 &    +0.024 & -3.377 $\pm$ 0.047 & -0.037 &  -3.414 $\pm$ 0.059 &  24  &  4.305  \\
LP\,412-31 & Ca2610c522 &    -0.165 & -3.246 $\pm$ 0.061 & +0.057 &  -3.189 $\pm$ 0.036 &  12  & 38.805  \\
\hline
  \multicolumn{8}{@{}l@{}}{\hbox to 0pt{\parbox{150mm}{\footnotesize
      \hbox{}
      \par\noindent
      \footnotemark[$*$] Specified model for $T_{\rm eff}$ and log\,$g$ in the second and third columns of table\,6, respectively.
      \par\noindent
      \footnotemark[$\dagger$] First correction to the starting value of log\,$A_{\rm C}$ = log\,$A_{\rm C}^{(1)}$ in table 6 
         and log\,$A_{\rm O}$ = log\,$A_{\rm C}^{(1)}$ + 0.30.
      \par\noindent
      \footnotemark[$\ddagger$] Second correction to the starting values log\,$A_{\rm C}$ = log\,$A_{\rm C}^{(2)}$ in the ninth column 
         of table\,6 and log\,$A_{\rm O}$ = log\,$A_{\rm O}^{(1)}$ in fourth column of this table.
      \par\noindent
      \footnotemark[$\S$]  Number of H$_2$O blends used for the mini-CG analysis. 
      \par\noindent
      \footnotemark[$\|$]  $\chi^{2}$ value for the comparison of the observed and  predicted spectra in figure\,11.
      }\hss}}
  \end{tabular}
\end{center}
\end{table*}

\begin{table*}
\begin{center}
\caption{log\,$(W/\lambda)_{\rm obs}$\footnotemark[$*$] of 
H$_2$O blends in region A for 7 late M dwarfs.}
  \begin{tabular}{cccccccc}
  \hline
 Ref. no\footnotemark[$\dagger$] & GJ\,54.1 & GJ\,752B & GJ\,873 & GJ\,1002 
& GJ\,1245B & GAT\,1370 & LP\,412-31  \\ 
\hline
    A01 &    -4.797 & -4.606 & -4.814 & -4.758 & -4.724 & -4.622 & -4.666 \\
    A02 &     ---   &  ---   & -4.791 & -4.696 &  ---   &  ---   &  ---   \\
    A03 &    -4.524 & -4.347 & -4.620 & -4.475 & -4.456 & -4.349 & -4.355  \\
    A04 &    -4.399 & -4.235 & -4.467 & -4.335 & -4.332 & -4.208 & -4.260  \\
    A05 &    -4.999 & -4.764 & -5.047 & -4.983 & -4.909 & -4.754 & -4.814  \\
    A06 &     ---   &  ---   & -4.832 & -4.699 &  ---   &  ---   &  ---    \\
    A07 &    -4.713 &  ---   & -4.722 & -4.617 &  ---   & -4.479 &  ---    \\
    A08 &    -4.830 & -4.605 & -4.910 & -4.764 & -4.744 & -4.648 & -4.585  \\
    A09 &    -4.761 &  ---   & -4.813 & -4.673 &  ---   &  ---   &  ---   \\ 
    A10 &     ---   &  ---   & -4.762 & -4.636 &  ---   &  ---   &  ---    \\
    A11 &    -4.570 & -4.408 &  ---   & -4.503 & -4.469 &  ---   & -4.411  \\
    A12 &     ---   &  ---   &  ---   &  ---   &  ---   &  ---   &  ---  　\\
    A13 &    -4.493 & -4.325 & -4.565 & -4.412 & -4.419 &  ---   &  ---    \\
    A14 &    -4.722 &  ---   & -4.799 & -4.659 &  ---   &  ---   &  ---    \\
    A15 &     ---   &  ---   &  ---   &  ---   &  ---   &  ---   &  ---    \\
    A16 &     ---   &  ---   &  ---   &  ---   &  ---   &  ---   &  ---    \\
    A17 &     ---   & -4.362 & -4.674 & -4.482 & -4.502 & -4.348 &  ---    \\
\hline
  \multicolumn{8}{@{}l@{}}{\hbox to 0pt{\parbox{150mm}{\footnotesize
      \hbox{}
      \par\noindent
      \footnotemark[$*$] The equivalent width $W$ is measured by referring to the 
        pseudo-continuum.
      \par\noindent
      \footnotemark[$\dagger$]  Ref. no. defined in table 2 and figure\,2 of Paper II.
      }\hss}}
  \end{tabular}
\end{center}
\end{table*}

\begin{table*}
\begin{center}
\caption{Oxygen abundances log\,$A_{\rm O}^{A}$ from the H$_2$O blends in region A.}
  \begin{tabular}{lcccrr}
  \hline
obj.  & model2\footnotemark[$*$] & $\Delta$\,log\,$A_{\rm O}^{(3)}$\footnotemark[$\dagger$]
 & log\,$A_{\rm O}^{(3)}$   & $N_{\rm A}$\footnotemark[$\ddagger$] &  $\chi^{2}$\footnotemark[$\S$] \\
\hline
GJ\,54.1   & Ca3160c508 & +0.032 & -3.111 $\pm$ 0.065 &  13 &   4.485 \\      
GJ\,752B   & Cc2640c521 & -0.046 & -3.347 $\pm$ 0.028 &   8 &  20.272 \\     
GJ\,873    & Ca3430c490 & -0.088 & -3.157 $\pm$ 0.032 &  15 &   7.313 \\     
GJ\,1002   & Ca2970c513 & -0.035 & -3.188 $\pm$ 0.063 &  17 &  25.032 \\     
GJ\,1245B  & Ca2940c514 & -0.041 & -3.173 $\pm$ 0.097 &  11 &   6.822 \\     
GAT\,1370  & Cc2690c520 & +0.003 & -3.411 $\pm$ 0.056 &   9 &  14.985 \\     
LP\,412-31 & Ca2610c522 & -0.007 & -3.207 $\pm$ 0.016 &   7 &   1.970 \\     
\hline
  \multicolumn{6}{@{}l@{}}{\hbox to 0pt{\parbox{120mm}{\footnotesize
      \hbox{}
      \par\noindent
      \footnotemark[$*$] Specified model for $T_{\rm eff}$ and log\,$g$ in the second and third columns of table\,6, respectively.
      \par\noindent
      \footnotemark[$\dagger$] Correction to the starting values log\,$A_{\rm C}$ = log\,$A_{\rm C}^{(2)}$ in the ninth column 
         of table\,6 and log\,$A_{\rm O}$ = log\,$A_{\rm O}^{(2)}$ in sixth column of  table\,8.
      \par\noindent
      \footnotemark[$\ddagger$]  Number of H$_2$O blends in region A used for the mini-CG analysis. 
      \par\noindent
      \footnotemark[$\S$]  $\chi^{2}$ value for the comparison of the observed and  predicted spectra in figure\,12.
      }\hss}}
  \end{tabular}
\end{center}
\end{table*}

\begin{table*}
\begin{center}
\caption{Mini curve-of-growth analysis on CO and H$_2$O lines in 2MASS 1835+32.}
  \begin{tabular}{clcllcccc}
  \hline
no. & model & $V_{\rm rot}{\rm sin}\,i$\footnotemark[$*$]  & log\,$A_{\rm C}^{(0)}$\footnotemark[$\dagger$] &  
log\,$A_{\rm O}^{(0)}$\footnotemark[$\ddagger$]  & $\Delta\,{\rm log}\,A_{\rm C}$\footnotemark[$\S$] &
$\Delta\,{\rm log}\,A_{\rm O}$\footnotemark[$\|$] & log\,$A_{\rm C}$\footnotemark[$\#$] &
log\,$A_{\rm O}$\footnotemark[$**$] \\
\hline
1 & Bc2280c526 & 40.8  &  -3.61  & -3.31   & -0.101 &           & -3.711 $\pm$ 0.128 &            \\
2 & Bc2280c526 & 40.8  &  -3.711 & -3.711+0.3 &           & +0.008  &           &  -3.403 $\pm$ 0.024 \\
3 & Bc2280c526 & 40.8  &  -3.711 & -3.403  & -0.025 &           & -3.736 $\pm$ 0.130 &             \\
4 & Bc2280c526 & 40.8  &  -3.736 & -3.403  &           & -0.012 &           &   -3.415 $\pm$ 0.023 \\
5 & Cc2280c526 & 40.8  &  -3.61  & -3.31   & -0.358 &           & -3.968 $\pm$ 0.078 &             \\
6 & Cc2280c526 & 40.8  &  -3.968 & -3.968+0.3 &           & +0.144 &           &  -3.524 $\pm$ 0.034  \\
\hline
7 & Bc2280c526 & 37.6  & -3.61   & -3.31   & -0.184 &           & -3.794 $\pm$ 0.094 &               \\
8 & Bc2280c526 & 37.6 &  -3.794 & -3.794+0.3 &           & +0.033 &           &  -3.461 $\pm$ 0.036     \\
9 & Bc2280c526 & 44.0  &  -3.61  & -3.31   & +0.044 &           & -3.566 $\pm$ 0.199 &              \\
10 & Bc2280c526 & 44.0  & -3.566  & -3.566+0.3 &           & -0.052 &           &  -3.318 $\pm$ 0.029   \\
\hline
11 & Bc2180c528 & 40.8  & -3.61  & -3.31   & +0.0003 &           & -3.610 $\pm$ 0.136 &                \\
12 & Bc2180c528 & 40.8  & -3.610 & -3.610+0.3 &           & -0.014    &           &   -3.324 $\pm$ 0.023   \\
13 & Bc2380c525 & 40.8  & -3.61  & -3.31   & -0.094 &           & -3.704 $\pm$ 0.142 &                \\
14 & Bc2380c525 & 40.8  & -3.704 & -3.704+0.3 &           & +0.016    &           &   -3.388 $\pm$ 0.023  \\
\hline
  \multicolumn{9}{@{}l@{}}{\hbox to 0pt{\parbox{150mm}{\footnotesize
      \hbox{}
      \par\noindent
      \footnotemark[$*$] In km\,s$^{-1}$. 
      \par\noindent
      \footnotemark[$\dagger$]  A starting value for log\,$A_{\rm C}$ in the mini CG analysis.
      \par\noindent
      \footnotemark[$\ddagger$]  A starting value for log\,$A_{\rm O}$ in the mini CG analysis.
      \par\noindent
      \footnotemark[$\S$]  Abundance correction to log\,$A_{\rm C}$ in column 4. 
      \par\noindent
      \footnotemark[$\|$]  Abundance correction to log\,$A_{\rm O}$ in column 5.  
      \par\noindent
      \footnotemark[$\#$]  Resulting abundance log\,$A_{\rm C}$ = log\,$A_{\rm C}^{(0)}$
            +$\Delta\,{\rm log}\,A_{\rm C}$. 
      \par\noindent
      \footnotemark[$**$]  Resulting abundance log\,$A_{\rm O}$ = log\,$A_{\rm O}^{(0)}$
            +$\Delta\,{\rm log}\,A_{\rm O}$.  
      }\hss}}
  \end{tabular}
\end{center}
\end{table*}

\begin{table*}
\begin{center}
\caption{$\chi^{2}$ test on the fittings of the observed and predicted spectra of CO and H$_2$O  
in 2MASS 1835+32.}
  \begin{tabular}{clcllcccc}
  \hline
no. & model & $V_{\rm rot}\,{\rm sin}\,i$\footnotemark[$*$]  & log\,$A_{\rm C}$\footnotemark[$\dagger$] &  
log\,$A_{\rm O}$\footnotemark[$\ddagger$]  & $\chi^{2}$(CO)\footnotemark[$\S$] & $N_{\rm CO}$ 
\footnotemark[$\|$]  & $\chi^{2}$\,(H$_2$O)\footnotemark[$\#$] & $N_{\rm H_{2}O}$\footnotemark[$**$] \\
\hline
1 & Bc2280c526 & 40.8 & -3.711 & -3.403 &   5.803 &     4 &  2.035 &  8  \\
2 & Cc2280c526 & 40.8 & -3.968 & -3.524 &  10.369 &     4 &  2.879 &  8  \\
\hline
3 & Bc2280c526 & 37.6 & -3.794 & -3.461 &   9.406 &     4 &  2.568 &  8   \\
4 & Bc2280c526 & 44.0 & -3.566 & -3.318 &   3.126 &     4 &  1.820 &  8   \\
\hline
5 & Bc2180c528 & 40.8 & -3.610 & -3.324 &   5.682 &     4 &  2.074 &  8    \\
6 & Bc2380c525 & 40.8 & -3.704 & -3.388 &   6.016 &     4 &  2.099 &  8    \\
\hline
  \multicolumn{9}{@{}l@{}}{\hbox to 0pt{\parbox{120mm}{\footnotesize
      \hbox{}
      \par\noindent
      \footnotemark[$*$] In km\,s$^{-1}$. 
      \par\noindent
      \footnotemark[$\dagger$]  log\,$A_{\rm C}$ used for predicted spectrum.
      \par\noindent
      \footnotemark[$\ddagger$]  log\,$A_{\rm O}$ used for predicted spectrum.
      \par\noindent
      \footnotemark[$\S$]  $\chi^{2}$\,value for the fitting of the observed
         and predicted spectra of CO.
      \par\noindent
      \footnotemark[$\|$]  Number of CO blends used in the mini CG analysis.
      \par\noindent
      \footnotemark[$\#$]  $\chi^{2}$\,value for the fitting of the observed
         and predicted spectra of H$_{2}$O in region B.   
      \par\noindent
      \footnotemark[$**$] Number of H$_{2}$O blends used in the mini CG analysis.
      }\hss}}
  \end{tabular}
\end{center}
\end{table*}

\begin{table}
\begin{center}
\caption{Carbon and oxygen abundances in late M dwarfs.}
  \begin{tabular}{lccc}
  \hline
obj.  & log\,$A_{\rm C}$\footnotemark[$*$]
 & log\,$A_{\rm O}$\footnotemark[$\dagger$]   & log\,$A_{\rm O}/A_{\rm C}$ \\
\hline
GJ\,54.1   & -3.38 $\pm$ 0.09  & -3.13 $\pm$ 0.06 & 0.25 \\
GJ\,752B   & -3.55 $\pm$ 0.07  & -3.31 $\pm$ 0.04 & 0.24 \\
GJ\,873    & -3.44 $\pm$ 0.09  & -3.13 $\pm$ 0.06 & 0.31 \\
GJ\,1002   & -3.42 $\pm$ 0.09  & -3.17 $\pm$ 0.05 & 0.25 \\
GJ\,1245B  & -3.44 $\pm$ 0.10  & -3.15 $\pm$ 0.07 & 0.29 \\
GAT\,1370  & -3.78 $\pm$ 0.06  & -3.41 $\pm$ 0.06 & 0.37 \\
LP\,412-31 & -3.36 $\pm$ 0.07  & -3.20 $\pm$ 0.03 & 0.16 \\
2MASS 1835+32 & -3.57 $\pm$ 0.20  & -3.32 $\pm$ 0.03 & 0.25 \\
the Sun 1\footnotemark[$\ddagger$] & -3.44 $\pm$ 0.04  & -3.07 $\pm$ 0.035 & 0.37 \\
the Sun 2\footnotemark[$\S$] & -3.57 $\pm$ 0.05  & -3.31 $\pm$ 0.05 & 0.26 \\
\hline
  \multicolumn{4}{@{}l@{}}{\hbox to 0pt{\parbox{80mm}{\footnotesize
      \hbox{}
      \par\noindent
      \footnotemark[$*$] From table 6 for the first seven objects and
         from table 11 (line no.\,9) for 2MASS 1835+32.
      \par\noindent
      \footnotemark[$\dagger$] The weighted mean of
          log\,$A_{\rm O}^{B}$ and log\,$A_{\rm O}^{A}$ from 
          table\,8 and table\,10, respectively, for the first seven objects
          and from table 11 (line no.\,10) for 2MASS 1835+32 for which region A is not
          analyzed. 
      \par\noindent
      \footnotemark[$\ddagger$]  \citet{And89}. 
      \par\noindent
      \footnotemark[$\S$]  \citet{Asp09}.
      }\hss}}
  \end{tabular}
\end{center}
\end{table}

\begin{table}
  \caption{ Spectroscopic data of CO lines (2-0 band).\footnotemark[$*$] }
  \begin{center}
    \begin{tabular}{ccccc}
      \hline
  Ref. no.\footnotemark[$\dagger$] & wavelength\footnotemark[$\ddagger$]
& log\,$gf$ & L.E.P. & Rot. ID.  \\
     &  (\AA)     &            & (cm$^{-1}$) &      \\
      \hline
   15 &  23029.797  &  -5.499 & 1667.971 &  R 29 \\
      &  23029.902  &  -4.984 & 9936.531 &  R 72 \\
\hline
\multicolumn{5}{l}{\hbox to 0pt{\parbox{80mm}{\footnotesize
   \par\noindent
   {}
   \par\noindent
     \footnotemark[$*$] To be added to table\,7 of Paper I.
     \par\noindent
     \footnotemark[$\dagger$] Ref.\,no. refers to figure 13b.
     \par\noindent
     \footnotemark[$\ddagger$] In vacuum.
     }}}
\end{tabular}
\end{center}
\end{table}

\end{document}